\documentclass[3p,times,12pt]{elsarticle}
\journal{arXiv}

 \usepackage{graphicx}
 \usepackage{amsmath}
 \usepackage{amsfonts}
 \usepackage{subcaption,graphicx}
 \usepackage{wrapfig}
 \usepackage{enumerate} 
 \usepackage{adjustbox}
 \makeatletter
 \usepackage{listings}
 \usepackage{color}
 \usepackage[dvipsnames]{xcolor}
 \usepackage{multicol}
 \usepackage{soul}  
 \usepackage{multicol}
\usepackage{enumitem}
\date{}

\biboptions{sort&compress}
\usepackage[T1]{fontenc}

\begin{document}

\newcommand{\JHS}[1]{\textcolor{black}{#1}}
\newcommand{\JHSS}[1]{\textcolor{black}{#1}}

\newenvironment{thisnote}{\par\color{black}}{\par}

\begin{frontmatter}
\title{Robust correlation measures for informative frequency band selection in heavy-tailed vibration signal}

\author[label1]{Justyna Hebda-Sobkowicz}
\author[label1]{Radosław Zimroz}
\author[label2]{Anil Kumar\corref{cor1}}
\author[label3]{Agnieszka Wyłomańska}

\address[label1]{Faculty of Geoengineering, Mining, and Geology, Wroclaw University of Science and Technology, \\Na Grobli 15, 50-421 Wroclaw, Poland, \{justyna.hebda-sobkowicz, radoslaw.zimroz\}@pwr.edu.pl\\}
\address[label2]{College of Mechanical and Electrical Engineering, Wenzhou University, Wenzhou, 325 035, China, 20210129@wzu.edu.cn}
\address[label3]{Faculty of Pure and Applied Mathematics, Hugo Steinhaus Center, Wroclaw University of Science and Technology, Wybrze{\.z}e Wyspia{\'n}skiego 27, 50-370 Wroclaw, Poland, agnieszka.wylomanska@pwr.edu.pl}

\begin{abstract}
Vibration signals are commonly used to detect local damage in rotating machinery. However, raw signals are often noisy, particularly in crusher machines, where the technological process (falling pieces of rock) generates random impulses that complicate detection. To address this, signal pre-filtering (extracting the informative frequency band from noise-affected signals) is necessary. This paper proposes an algorithm for processing vibration signals from a bearing used in an ore crusher. Selecting informative frequency bands (IFBs) in the presence of impulsive noise is notably challenging. The approach employs correlation maps to detect cyclic behavior within specific frequency bands in the time-frequency domain (spectrogram), enabling the identification of IFBs. Robust correlation measures and median filtering are applied to enhance the correlation maps and improve the final IFB selection. Signal segmentation and the use of averaged results for IFB selection are also highlighted. The proposed trimmed and quadrant correlations are compablack with the Pearson and Kendall correlations using simulated \JHSS{signal, real vibration signal from crusher in mining industry and acoustic signal measublack on the test rig. Furthermore,} the results of real vibration analyses are compablack with established IFB selectors, including the spectral kurtosis, the alpha selector and the conditional variance-based selector.

\end{abstract}


\end{frontmatter}
\section*{Nomenclature}
\addcontentsline{toc}{section}{List of Abbreviations}

\begin{description}[labelwidth=1.3cm, leftmargin=3.5cm, labelsep=0.5cm,  font=\normalfont]
  \item[ACI] Amplitude of cyclic impulses
\item[ANCI] Amplitude of noncyclic impulses
\item[CM] Correlation map
\item[ENVSI] Envelope spectrum based indicator
\item[IFB] Informative frequency band
\item[KCC] Kendall correlation coefficient
\item[PCC] Pearson correlation coefficient
\item[QCC] Quadrant correlation coefficient
\item[SNR] Signal to noise ratio
\item[SOI] Signal of interest
\item[TCC] Trimmed correlation coefficient
	\end{description}

\section{Introduction}
Local damage detection in gearboxes or rolling element bearings is based primarily on the processing of vibration signals \cite{antoni2002differential,randall2011rolling,feng2023review}. Local damage produces cyclic and impulsive signals, and basic formulas allow one to calculate the so-called fault frequencies and then detect these components in the envelope spectrum. If such a signature can be found, maintenance actions should be taken. However, in practice, damage detection is challenging, especially if the signal of interest (SOI, i.e., the impulsive and cyclic component) is weak, noise is covering the SOI (resulting in a poor signal-to-noise ratio, SNR), the cycle length (i.e., the period between impulses) is time-varying, and the SOI is masked by non-Gaussian (often impulsive, heavy-tailed) noise. Heavy-tailed distributions are probability distributions whose tails are not exponentially bounded, which means that they have tails heavier than the exponential distribution \cite{asmussen2003steady}. Non-informative components with impulsive character could be linked to various technological procedures such as crushing, milling, screening, cutting, compressing, etc. \cite{Cocconcelli2012667,Hebda-Sobkowicz2020,Kruczek2020,Schmidt2020}, other faults in complex mechanical systems \cite{Xia2018,wodecki2020separation,Liu2024,Huo2024} or imperfection of measurement systems (that is, electromagnetic interference \cite{mauricio2020bearing}).

The SOI can be interpreted as an amplitude-modulated signal with the fault frequency (and harmonics) as the modulating signal and the structural resonance (natural) frequency as a carrier. In the time-frequency domain, SOI is a cyclic wideband excitation in some frequency bands. It allows for the selection of IFBs where the SOI exists. The most popular approach is envelope analysis and detection of fault frequencies in the envelope spectrum (ES) of a prefilteblack signal. Enhanced versions of the envelope are also known, such as squablack envelope spectrum (SES) \cite{randall2001relationship,borghesani2016cyclostationary}, log-envelope spectrum (LES)  \cite{borghesani2017cs2,smith2019optimal} or product envelope spectrum (PES) \cite{Chen2023}. Its comparison can be found in \cite{Chen2023}. Many methods have been developed for vibration prefiltering \cite{ni2019rolling,Schmidt_Sensors,mauricio2020improved,Alavi2022,Yu2019375,Wang20222762,Wang2022mst,Wang2022,Dragomiretskiy2014531,Lu2021,Yang2023,Liu2024mst} to extract SOI (or at least improve SNR).  

Significant advances have been made in automated fault diagnosis of rotating machinery, particularly through the application of advanced machine learning techniques \cite{tang2021improved,tang2022novel,yan2022semi,wang2023adaptive,wu2023conditional,dong2024multi,xu2024multi,lai2024automated}. In \cite{lai2024automated}, the authors proposed a subdomain greedy network architecture search for fault diagnosis, while a semi-supervised method robust to speed fluctuations and low labeled data rates was introduced in \cite{yan2022semi}. Moreover, multisensor approaches have been developed for more reliable online fault detection \cite{xu2024multi, dong2024multi}. In \cite{tang2022novel}, the authors focused on adaptive convolution neural networks to diagnose hydraulic piston pump faults using acoustic images, further improving the approach with adaptable learning rates for multi-signal diagnosis \cite{tang2021improved}. These advances highlight the growing complexity and precision of fault diagnosis methods. However, challenges remain in balancing performance with simplicity.

Many approaches are related to testing the impulsiveness or periodicity of the signal in the time-frequency domain \cite{antoni2007fast,combet2009optimal,antoni2016info,li2016extracting,zak20152987,wang2016extension,hebda2022infogram}. The results are then projected onto the frequency axis to obtain an IFB selector as a function of frequency. The most well-known criterion for detecting impulsiveness is kurtosis \cite{randall2011rolling,combet2009optimal}. If kurtosis is applied to the family of subsignals obtained by the filter bank, then the distribution of measures of impulsiveness versus frequencies can be obtained. This method has been used within the framework of kurtogram \cite{antoni2007fast} or spectral kurtosis \cite{antoni_randall,combet2009optimal}. However, its main limitation is the inability to recognize whether the impulses that occur are cyclic or not. In addition, kurtosis is very sensitive to impulsive noise.  Therefore, there are other extensions that aim to minimize these drawbacks \cite{antoni2016info,wang2018some,miao2017improvement_GINI,li2016extracting,wang2016extension,wang2020sum,wang2016new,liu2019accugram,ni2021novel}. One of them considers the utilization of the $\alpha$-stable distribution, which covers a wide class of distributions, from the classical Gaussian distribution to the strong heavy-tailed Levy distribution. The stability parameter $\alpha$ of the $\alpha$-stable distribution was proposed as the IFB selector \cite{zak20152987} (alpha selector), which performs filtration much better than popular spectral kurtosis, especially in signals with heavy-tailed noise (more resistant to outliers). Another approach that also aims to minimize the influence of random noncyclic impulses on background noise is the use of a conditional variance-based statistic for IFB selection, known as a conditional variance-based selector (CVB selector) \cite{Hebda-Sobkowicz2020}. The advantage of this selector is that it responds to impulsivity (similar to the spectral kurtosis and alpha selector mentioned above), while blackucing the influence of outliers by trimming the tails of the distribution.
Comparison of recently developed methods and new methods for IFB selection in the presence of non-Gaussian noise is available in \cite{liu2015bearing,wodecki2019novel,Zhou2020,Hebda-Sobkowicz2020_AS,nowicki2021dependency,hebda2021alternative,hebda2022infogram}.
However, the problem of signal prefiltering remains a challenging issue \cite{Kim2022}. 

In this paper, robust correlation measures are presented, that is, quadrant and trimmed correlation coefficients (QCC and TCC, respectively), to search for information related to local fault in the time-frequency domain. 
This work could be consideblack as an extension of \cite{nowicki2021dependency}. In \cite{nowicki2021dependency} authors have presented the advantage of applying the Kendall correlation coefficient and the Spearman correlation coefficient to the correlation map (CM). The final IFB selectors based on Kendall and Spearman correlation coefficients have similar results for the consideblack signals and seem to be especially useful in the case of heavy-tailed data. The results were compablack with the classical correlation measure, that is, the Pearson correlation coefficient (PCC).
In this paper, one of them, the Kendall correlation coefficient (KCC), was chosen as the robust approach during the comparison of the new proposed robust correlation measures.  The results of PCC are also presented as the most well-known and simplest method.

The CM is a symmetric matrix of correlation values calculated for the pairs of frequency bins of the spectrogram. If the subsignal in the frequency band $f_i$ exhibits a behavior similar to that of the
subsignal in the frequency band $f_j$, it means that it should be treated as the same source of information. High correlation values are expected for frequency bands that contain cyclic impulsive components (related to local damage). The more robust the measure of correlation is, the more zero the correlation values are for bands containing only noise and outliers (without the cyclic properties). The information sought on the cyclic behaviors is then clear.
However, the method presented in \cite{nowicki2021dependency} has some drawbacks. It has appeablack that even if KCC and SCC are robust correlation measures, some non-informative components are still present in the correlation map. This is a consequence of the existence of a non-zero correlation between noise with non-cyclic high-energy impulses that are broadband in the frequency domain. Then, for the frequency bands tested, $f_i$ and $f_j$, which are close enough to each other, these broadband impulses appear in the same timestamp, making the correlation higher. In \cite{nowicki2021dependency} authors proposed the aggregation of information from the two-dimensional (2D) CM to the one-dimensional (1D) vector (CM-based IFB selector) used as the filter characteristic. During aggregation, the imperfection of the CM is even more highlighted. Therefore, a threshold has been proposed to improve its selectivity. However, the proposed threshold is not universal and a new procedure is needed. Moreover, the computational cost of KCC is very high. It was the motivation to modify this procedure. 

In this paper, the novel application of robust correlation measures, i.e. TCC and QCC in CM, is presented. More information on robust correlation measures can be found in \cite{croux2010influence,durre2015robust,Żuławiński2024}. 
The effectiveness of proposed robust correlation measures is visually checked by comparison of CMs and CM-based IFB selectors based on the proposed robust correlation measures (TCC and QCC) and commonly known PCC and KCC. To improve the identification of the cyclic component, enhancement of the CM is proposed, i.e. the median filter, which can remove random spikes in the map. Median filtering is a non-linear operation often used in image processing to blackuce "salt and pepper" noise \cite{lim1990two}. In the paper, a 2D median filter is applied to the CM. Subsequently, a new CM (with a median filter applied) contains the median value in a 3-by-3 neighborhood around the corresponding value of the original CM. Therefore, the randomly occurring high correlation values for single bins of frequencies are minimized. In addition, splitting the signal into shorter length signals and averaging the result of all segments is also discussed in this paper as an idea to significantly smooth the results of the CM-based IFB selector. It seems to be especially useful in the case of signals with non-cyclic impulses.

 The paper considers the following novel enhancement of the correlation maps and correlation map-based IFB selectors as follows:
 \begin{itemize}
     \item propose new measures for correlation estimation which are less sensitive for outlier values comparing to known Pearson correlation or even Kendall correlation
     \item propose measures which have significantly lower computational costs than known Kendall correlation measure 
     \item propose new correlation map enhancement which are automatic and universal in case of some outlier values appearance and can be applied for any correlation maps
     \item propose automatic and universal correlation map-based IFB selector enhancement which can be applied for any IFB selectors.
 \end{itemize}

Calculation of the correlation map, map cleaning by median filter averaging, and signal segmentation to obtain the averaged results) allow one to estimate a novel effective selector, useful in SOI extraction in the presence of impulsive noise. The small value of the selector for a given frequency bin means that this band does not contain SOI, whereas larger values indicate more information about the consideblack frequency ranges. The output of the procedure is a vector of selector values for all frequencies (similarly to spectral kurtosis).
Finally, the values of the envelope spectrum-based indicator (ENVSI) \cite{Hebda-Sobkowicz2020} and the computational costs of the trimmed, quadrant, Kendall, and Pearson CMs are compablack for the simulation signals. 
The proposed procedure has also been compablack to classical spectral kurtosis and recently developed selectors suitable for non-Gaussian background noise, namely the alpha selector and CVB selector \cite{Hebda-Sobkowicz2020_AS,Hebda-Sobkowicz2020}.
The methodology is also tested on the real vibration signal of the bearing of the crusher machine. 

The rest of the paper is organized as follows. Section~\ref{metodo} presents the basic definitions of the correlation measures used, and the CM structure is referenced. Section~\ref{new_proposed} contains a description of the proposed approach and the step-by-step procedure of the algorithm. In Section~\ref{model}, the model of the simulation data is presented. \JHSS{In Section ~\ref{results}, the proposed methodology is examined using simulation data with different SNRs, by varying the SOI amplitude versus constant parameters of Gaussian noise and by changing the amplitude of non-cyclic impulses. The proposed approach is also tested on real vibration data from a hammer crusher operating in the mining industry, as well as acoustic data from the bearing of a test rig electric motor. Comparisons with known methods, i.e., spectral kurtosis, the alpha selector, and the CVB selector, are also provided. Section~\ref{conc} summarizes the results.} 

\section{Basic definitions}
\label{metodo}

 In this paper, the raw vibration signal $\mathbf{x}$ is transformed into the time-frequency domain by STFT. The following definition of the STFT is used:
\begin{equation}
    STFT_x(f,t)=\sum_{h=1}^{N}x_h w_{t-h}e^{\frac{-2j\pi fh}{N}},
    \label{eq:stft}
\end{equation}
where $w_{t-h}$ is the shift window, $x=(x_1,\dots,x_N)$ is the input signal of length $N$, $t=t_1,\dots,t_T$ is the time point, $f=f_1,\dots,f_F$ is a frequency, and $j$ is an imaginary unit; see \cite{boash} for more details. The spectrogram means the absolute values of the STFT, that is, $spec(t,f)=|STFT(t,f)|$.
During data analysis, the commonly used Hamming window of length 256, with 217 overlapping samples, was applied, and the Fourier transform was calculated for 512 frequency points. 

The subsignals of the spectrogram, that is, $subs(i)=spec(t,f_i)$ that correspond to the frequency bins $f_i$, are used to test the correlation. The symmetric correlation map (CM) is created, i.e.:
 \begin{equation}
\label{eq:sab}
CM(f_i,f_k)=corr(subs(i),subs(k)), \text{ 
 for } i,k=1,\dots,F.
 \end{equation}
 The function $corr(\cdot)$ denotes the empirical correlation. 
 The subsignals $subs(i)$ and $subs(k)$ show a similar behavior, which means that they should be treated as the same source of information. 

To investigate the existence of correlation in the time-frequency representation of the signal, the calculation of the correlation is performed by different correlation measures because some measures are sensitive to outliers. In this paper, four different correlation measures are investigated, that is, TCC, QCC, KCC, and PCC.

\subsection{Pearson correlation coefficient}

The PCC is the most known measure of correlation. It is used as a parametric correlation measure (the distribution of data with finite moments is assumed). PCC investigates the linear relationship within the data. The values of this PCC are in the range $[-1,1]$ and for values close to $-1$ and $1$ the strong linear relationship in the data should be observed. A value equal to 0 indicates that there is no correlation. 

Let $(x,y) = \{(x_1,y_1), (x_2,y_2), \dots, (x_N,y_N)\}$ be a two-dimensional sample of the random vector $(X,Y)$. The formula for the Pearson correlation coefficient of the sample vector $(x, y)$, denoted PCC, is defined as follows \cite{durre2015robust}:
\begin{equation}
\widehat{PCC}_{xy} =\frac{ \sum_{ i=1 }^{ N } (x_i - \overline{x})(y_i - \overline{y}) }{ \sqrt{ \sum_{ i=1 }^{ N } (x_i - \overline{x})^2} \sqrt{ \sum_{ i=1 }^{ N } (y_i - \overline{y})^2} },
\label{pearson1}
\end{equation}
where $\overline{x} $, $ \overline{y} $ are sample means of data vectors $x=(x_1,\dots,x_N)$ and $y=(y_1,\dots,y_N)$, respectively.
  The PCC estimator is sensitive to outliers. This means that the use of this correlation measure for data with a heavy-tailed distribution can give misleading results. Therefore, despite its simple definition and easy application, one can search for other correlation measures which could deal with heavy-tailed data.

  When the PCC is used as the measure of correlation in the CM, the originating CM is called Pearson's correlation map (Pearson CM).

\subsection {Kendall correlation coefficient}
The KCC is used as a nonparametric correlation measure known to be more resistant to outliers than the PCC \cite{mottonen1999robust}.
Let $(x,y) = \{(x_1,y_1), (x_2,y_2), \dots, (x_N,y_N)\}$ be a two-dimensional random sample from the random vector distribution $(X, Y)$. Then, the formula for the Kendall correlation coefficient of the sample vector $(x,y)$, denoted as KCC, can be written as follows \cite{kendall1938new}:
\begin{equation}\label{kendal}
   \widehat{KCC}_{xy} =\frac{ 2 }{ N(N-1) } \sum_{i=1}^{N-1}\sum_{j=i+1}^{N} J\Big((x_i,y_i),(x_j,y_j)\Big),
\end{equation}
where $J((x_i,y_i),(x_j,y_j)) =\text{sgn}(x_i-x_j) \text{sgn}(y_i-y_j)$ and function sgn$(\cdot)$ becomes $-1,0$ or $1$, if its argument is smaller, equal or greater than 0, respectively. The function $J(\cdot,\cdot)= 1,$ if a pair $(x_i,y_i)$ is concordant with a pair $(x_j,y_j)$ (have the same sign), that is, if $(x_i-x_j)(y_i-y_j) > 0$, while $J(\cdot,\cdot) =-1,$ if a pair $(x_i,y_i)$ is discordant with a pair $(x_j,y_j),$ i.e. if $(x_i-x_j)(y_i-y_j) < 0.$

The KCC is based on the difference between the probability that two variables are in the same order (for the observed data vector) and the probability that their order is different. The formula proposed by Kendall requires properly ordeblack data (each observed value is replaced by its number in an ascending sorted set) \cite{kendall1938new}. The KCC takes values between $[-1,1]$. The value equal to $1$ means complete accordance, the value equal to $0$ does not match orders, while the value equal to $-1$ means complete opposite accordance. The KCC indicates not only the strength but also the direction of the correlation. 

  When the KCC is used as the correlation measure in the CM, the originating CM is called Kendall's correlation map (Kendall CM).
\subsection{Quadrant correlation coefficient}

Let $(x,y) = \{(x_1,y_1), (x_2,y_2), \dots, (x_N,y_N)\}$ be a two-dimensional random sample from the distribution of the random vector $(X, Y)$.
The quadrant correlation coefficient of the random sample $(x,y)$, denoted as QCC, can be defined as follows \cite{durre2015robust}:
\begin{equation}
    \widehat{QCC}_{xy} = \frac{1}{N} \sum_{i=1}^{N}sgn\Big(\big(x_{i}-\tilde{x})(y_{i}-\tilde{y}\big)\Big),
    \label{eq:quadrant}
\end{equation}
where $\tilde{x}, \tilde{y}$ are median values. 
The QCC is calculated by first normalizing the data by the coordinate median. 
The QCC is equal to $1$ for a strong positive correlation, $-1$ for strong negative correlation, and $0$ for no correlation \cite{durre2015robust}.

When the sample QCC is used as the correlation measure in the CM, the originating CM is called the quadrant correlation map (quadrant CM).

\subsection {Trimmed correlation coefficient}
Let $(x,y) = \{(x_1,y_1), (x_2,y_2), \dots, (x_N,y_N)\}$ be a two-dimensional random sample from the distribution of the random vector $(X, Y)$ with finite variance. Let $z=(z_1,z_2,\dots,z_N)=(x_1y_1,x_2y_2,\dots,x_Ny_N)$ and $L$ be a $0-1$ function defined as:
\begin{equation}
L_i=
\begin{cases}
      1, & \text{if $z^{(k)}<z_i<z^{(N-k+1)}$},  \\
      0, & \text{otherwise,}
    \end{cases}  
\end{equation}
where $k=\lfloor c\cdot N\rfloor $ for $0\leq c < 0.5$, $c$ is a trimming constant or fraction of data in the sample $z$ of the pblackicted number of outliers and $z^{(j)}$ is an ordeblack vector in ascending order. The trimmed data vectors, assumed as devoid of outliers with a finite variance, fundamental for the calculation of TCC, are defined as follows:
\begin{align}
z_x^c=(x_1L_1,x_2L_2,\dots,x_NL_N),\label{def:trim1}\\
z_y^c=(y_1L_1,y_2L_2,\dots,y_NL_N).
\label{def:trim2}
\end{align}
The trimmed correlation coefficient of $(x,y)$, called TCC, can be defined as a PCC calculated for specially prepablack data included in sample vectors $z_x^c$ and $z_y^c$ defined in Eq.~(\ref{def:trim1}) and Eq.~(\ref{def:trim2}), i.e.: \begin{align}\widehat{TCC}_{xy}=\widehat{PCC}_{z_x^c,z_y^c}.\end{align}

TCC is a correlation measure based on sample Pearson correlation defined in Eq.~(\ref{pearson1}) but ignores the most extreme observations. The crucial point is the choice of \textcolor{black}{the parameter }$c$, which determines the final robustness of the TCC.  Higher values of $c$ can increase the robustness but decrease the efficiency for samples without
outliers. However, lower values might not trim out enough extreme values. In \cite{durre2015robust} authors recommend
using $c=3-5\%$ for medium-contaminated time series (in this paper, $c=3\%$ is established). This value was experimentally selected on the basis of the length of the signal, the sample frequency, and the parameters used during the calculation of the spectrogram. For the one-second signal with a sampling frequency of 25 kHz and the parameters used in Eq.~(\ref{eq:stft}) the spectrogram is the matrix with 256 samples of frequencies and 635 samples of time, which is a basis for the calculation of the CM.  Assuming 30 Hz as the damage frequency, one can conclude that the parameter $c$ should be lower than $c=\frac{30}{635}\approx 5\%$ to not remove the impulses of the fault. In practice, the fault impulses are not one-sample time series (which is described in detail in Section \ref{model}) and spread over a wider range of the timeline, so this proportion is always greater. However, as mentioned above, higher values of the trimmed parameter $c$ can improve robustness but may blackuce efficiency for samples that do not contain outliers. Other versions of the TCC statistic can also be found in \cite{durre2015robust}. As before, when the TCC is applied as the correlation measure in the CM map, the CM map that originated is called trimmed CM.

\section{\JHS{Proposed approach description}}
\label{new_proposed}
This article is a significant extension of recently published work \cite{nowicki2021dependency}, in which CM was introduced. The novelty part of this paper includes new REs application to the CM; median filter utilization for the map enhancement; averaged (median) IFB selector application for IFB selector improvement; testing the effectiveness of TCC, QCC, KCC, and PCC for variable amplitude parameters of cyclic and non-cyclic impulses and comparison of computational costs of the CMs calculated by different measures of correlation. The results obtained are more effective (more selective, automated, and faster) than the results presented in \cite{nowicki2021dependency} for the signal with non-Gaussian impulsive noise. 

\begin{figure}[ht!]
\centering
\includegraphics[width=\textwidth]{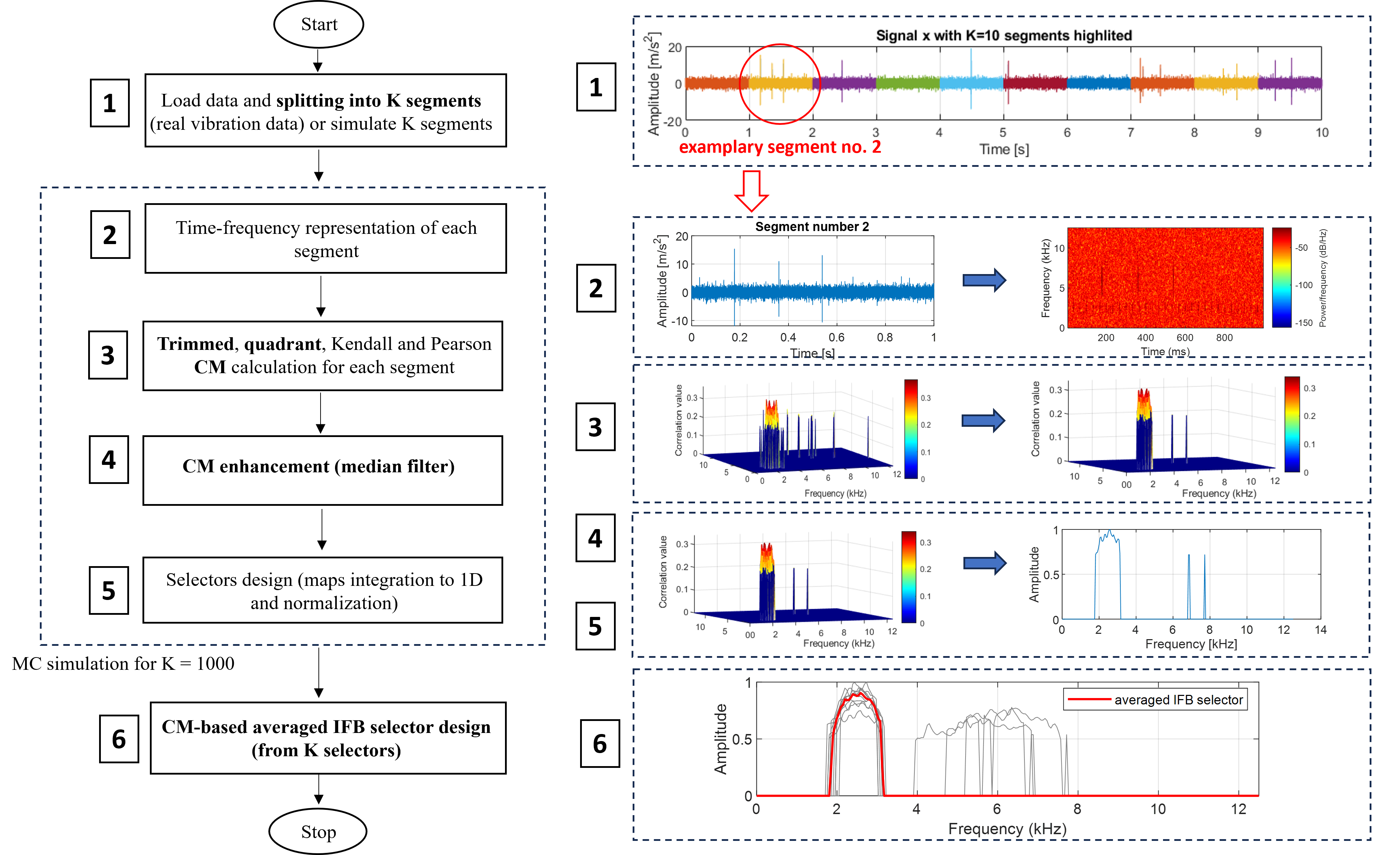}
\caption{\textcolor{black}{Block diagram and graphical illustration of the proposed data analysis procedure with the novelty part marked as a bold text.}}
\label{fig:block_diag}
\end{figure}

Fig.~\ref{fig:block_diag} (left panel) presents the schematic idea of the methodology with the novelty part marked in bold text.
The proposed methodology includes the following steps:
\begin{enumerate}[label=\arabic*.]
    \item Load data and split the signal into K shorter segments. 
The number K is chosen as a compromise between the length of the segment (to obtain optimal signal resolution) and the maximum number of segments. For the simulated signal, K=1000 was assumed. Graphic illustrations of each step of the methodology are presented in Fig.~\ref{fig:block_diag} (right panel) numbeblack according to the actual step of the methodology. 
    \item Transform a given segment into the time-frequency domain using STFT, described in Eq.~(\ref{eq:stft}). 
    \item Calculate the CM, see Eq.~(\ref{eq:sab}), using measure of correlation defined in Eq.~(\ref{pearson1}) -- (\ref{def:trim2}).
According to the methodology described in \cite{nowicki2021dependency} additional steps have to be performed to obtain enhanced CM, that is, remove the maximum values on the diagonal of the matrix and remove poor correlations. In \cite{nowicki2021dependency} authors proposed to use the third quartile. It was experimentally assumed value.
    \item Apply the median filter (analogously to image processing) to the matrix of correlations in order to smooth outliers (high value of correlation occurring in single frequency bands). The proposed median replaces the original value with the median of itself and 3 adjacent values.
    \item Aggregate the 2D map to a 1D IFB selector using the appropriate summation \cite{nowicki2021dependency} i.e. the conditional sum, which is the average amount of the nonzero values. Then, the IFB selector is obtained, which enables data filtration to extract SOI.
    \item Calculate the CM-based averaged IFB selector as the median of the selectors obtained from Steps 2 - 5.
\end{enumerate}
In \cite{nowicki2021dependency} the selector enhancement (thresholding) is also performed, during Step 5, to take into account only the highest selector values and to omit the selector values that correspond to the noise.
However, selecting the universal threshold is a challenge. The threshold for IFB selector described in \cite{nowicki2021dependency} is not universal,
arbitrarily added, therefore this part of the selector enhancement is omitted.
In this paper, improving the CM before aggregating the map into the 1D vector is proposed using the median filter. Then, such aggregation integrates both the information and the noise components (smoothed by the median filter). As we are dealing with randomly occurring impulses, we propose a fusion of information coming from several segments of the signal. It could be a longer signal divided into several segments (in the case of real vibration data) that correspond to several short simulated signals. The approach of splitting the data vector into shorter segments and considering the information from the averaged IFB selector calculated from the CMs of each segment is proposed as a universal enhancement of the IFB selector. It is assumed that the size of each segment includes at least 10 cycles of the local fault. 
For each segment, the correlation map is calculated, including map enhancement, and IFB selectors are designed. The final enhanced selector, see Step 6 in Fig.~\ref{fig:block_diag}, is the averaged value (median is used instead of classical arithmetic mean, as it is more resistant to the outliers value) of all obtained selectors corresponding to each segment, that is, the CM-based averaged IFB selector (referblack to as an averaged IFB selector).
The IFB selector is normalized using its maximum values to obtain the range $0-1$.
The normalized IFB selector is used for data filtration as a filter characteristic. Depending on the measure of the correlation used, the following selectors of the IFB are consideblack: Pearson IFB selector, Kendall IFB selector, trimmed IFB selector, and quadrant IFB selector. Their performance is tested using 1000 Monte Carlo (MC) simulations, i.e. Steps 1 -- 5 are repeated 1000 times for the simulated one-second heavy-tailed signal with local fault, described in the next section. MC simulations are also performed to change the parameters of the simulated signal. In case of real vibration data analysis, Steps 1 -- 5 are repeated for K segments, the number of which depends on the length of the signal and the sampling frequency.

To compare the efficiency of the filtering procedure, the ENVSI (envelope spectrum-based indicator) is used \cite{Hebda-Sobkowicz2020}.
It uses the squablack envelope spectrum (SES) of the filteblack signal to compare the portion of the energy related to fault frequency and it harmonics, i.e. amplitudes of impulses - AIS, 
to the whole energy of the squablack envelope spectrum. The higher the index value, the better the filtering result of a tested selector.
The consideblack indicator is defined as \cite{Hebda-Sobkowicz2020}: 
\begin{equation}
    ENVSI=\frac{\sum_{i=1}^{M}{\text{AIS}}}{\sum_{k=1}^{P}\text{
    SES}},
\end{equation}
where $M$ is the number of components to analyze corresponds to the impulses of fault frequency and $P$ is the number of frequency bins used to calculate the total energy. Note that the last component ($M$-th) of AIS is the same component ($P$-th) of SES. The number of harmonics used for the ENVSI calculation is arbitrarily chosen (set at 10), based on the total shape of the SES. 

The lack of an impulsive component in the consideblack SES implies that ENVSI converges to zero. When impulses appear in the time domain, informative components are present in the envelope spectrum and the ENVSI value increases. Depending on the level of the background noise, the value of ENVSI might be higher (low noise, i.e. visible impulses in the time domain)  or lower (barely visible impulses due to a high level of noise).
ENVSI makes it possible to compare the effectiveness of different diagnostic methods analyzed for the same signal or signals with similar noise complexity.

\section{Model of the \JHSS{simulated} signal}
\label{model}

The methodology described in Section~\ref{metodo}, for the calculation of the CM and IFB selection, was applied to the simulated signal $\mathbf{x}$. The signal $\mathbf{x}$ is inspiblack by real signals derived from the bearing vibration signals of the copper ore crusher, described in Section \ref{sec:exp}, where the technological process of the working machine generates many non-informative impulses that could be consideblack as heavy-tailed non-Gaussian background noise. 
A detailed description of the signal simulation is presented in \cite{Hebda-Sobkowicz2020}. The signal consists of three main components: $\mathbf{x_G}$, $\mathbf{x_{SOI}}$, and $\mathbf{x_{NC}}$, which form an additive mixture: $\mathbf{x}=\mathbf{x_G}+\mathbf{x_{NC}}+\mathbf{x_{SOI}}$. These components are as follows:
\begin{itemize}
    \item signal $\mathbf{x_G}\sim \mathcal{N}(\mu=0, \sigma=1)$, which is a Gaussian white noise,
    \item signal $\mathbf{x_{NC}}$ represents non-cyclic impulses related to the specificity of the machine operation (the localization and amplitudes of the impulses have the random, uniform distribution),
    \item and $\mathbf{x_{SOI}}$ is a signal of cyclic impulses associated with the local fault (amplitude of impulses is constant and location of the impulses is cyclic). 
\end{itemize}

\begin{figure}[ht!]
  \centering
\includegraphics[width=0.8\textwidth]{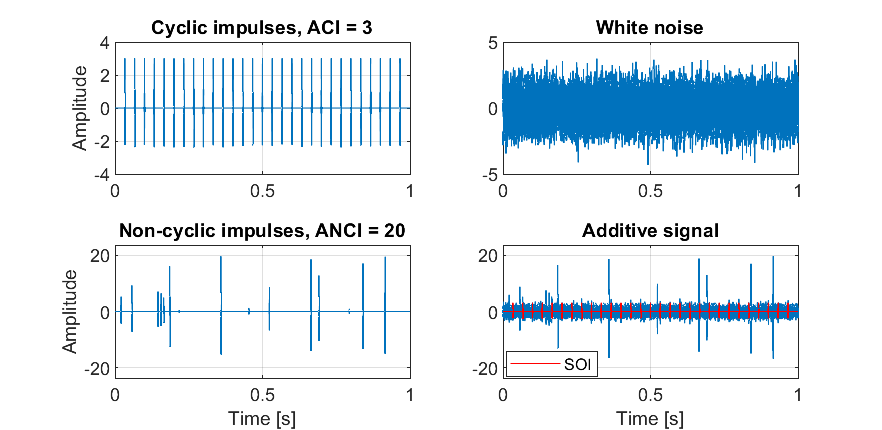}
    \caption{\textcolor{black}{Simulated additive signal $\mathbf{x}$ and its components.}}
    \label{fig:additive_sig}
\end{figure}

\begin{figure}[ht!]
  \centering
\includegraphics[width=0.73\textwidth]{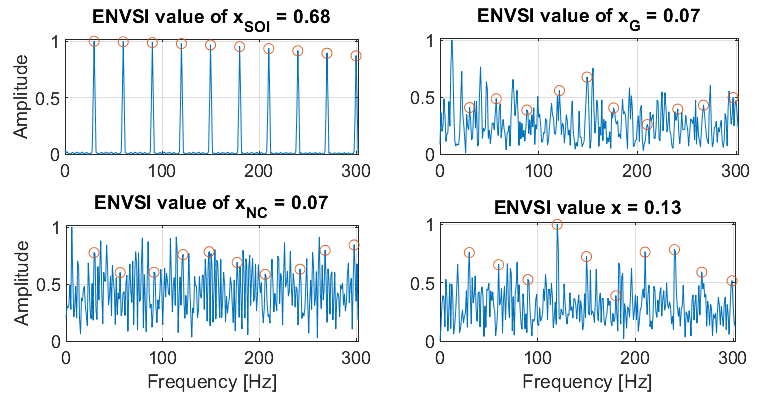}
    \caption{\textcolor{black}{The SES of the signal $\mathbf{x}$ and SES of its separated components with the frequency of fault and its harmonics marked in black circles.}}
    \label{fig:additive_envsi}
\end{figure}

\begin{figure}[ht!]
\begin{subfigure}[b]{0.48\textwidth}
\centering
\includegraphics[width=0.99\textwidth]{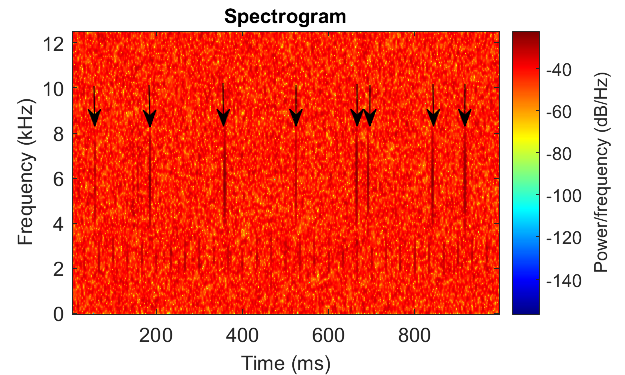}
\caption{}
\end{subfigure}\quad
\begin{subfigure}[b]{0.48\textwidth}
\centering
\includegraphics[width=0.99\textwidth]{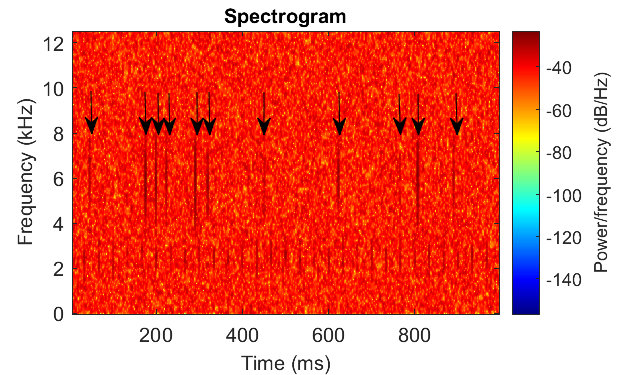}
\caption{}
\end{subfigure}
\caption{\textcolor{black}{The spectrograms of the simulated signals $\mathbf{x}$ (a) and $x'$ (b).}}
    \label{fig:meto_syg_spec}
\end{figure}

In the consideblack signal, non-cyclic impulses have random amplitudes (ANCI) according to the uniform distribution on the interval $[0,20]$: ANCI $\sim \mathcal{U}(0,20)$. They are generated using a Gaussian modulated sinusoidal pulse of unit amplitude at the times indicated in the vector $t_{nc}$, with a center frequency $cf_{nc}$ and a fractional bandwidth $bw_{nc}$. The distribution of the locations of the non-cyclic impulses (array $t_{nc}$) is uniform over the entire time interval. The amount of non-cyclic impulses in the signal of 1 second length is set at 15. The $cf_{nc}$ is set to 6 kHz. It corresponds to the natural frequency of the machine, a structural resonance. The fractional bandwidth is chosen from the uniform distribution in the interval $[0.4,0.5]$. 
The cyclic impulses follow the same steps of simulations as the non-cyclic impulses, but the dispersion of the locations of the impulses $t_{soi}$ corresponds to 30 Hz, and the amplitude of the cyclic impulses (ACI) is constant.
The central carrier frequency of the cyclic impulses ($cf_{soi}$) is set at 2.5 kHz (a bearing structural resonance). 
Assuming the parameters above, the final signal contains almost fully hidden cyclic impulses of the local fault, not visible in the time domain of the signal. Moreover, non-cyclic impulses additionally disturb the local fault detection procedure. 
Consequently, since the signal $\mathbf{x}$ is an additive mixture of white noise $\mathbf{x_G}$, a cyclic and impulsive component of fault $\mathbf{x_{SOI}}$ and non-cyclic impulses $\mathbf{x_{NC}}$, one can conclude that $\mathbf{x}$ is the heavy-tailed signal. Similar approaches can be found in \cite{Hebda-Sobkowicz2020,Hebda-Sobkowicz2020_AS}. The exemplary simulation of signal $\mathbf{x}$ and its separate components is presented in Fig.~\ref{fig:additive_sig}.
The SES of the signal and the SES of its separated components are presented in Fig.~\ref{fig:additive_envsi}.
As can be seen, the SOI (signal $\mathbf{x_{SOI}}$) is hidden in the background noise of the additive signal $\mathbf{x}$ (see Fig.~\ref{fig:additive_sig}), and the SES (see Fig.~\ref{fig:additive_envsi}) does not inform about the frequency of the SOI, as it is disturbed by the background heavy-tailed noise (i.e. $\mathbf{x_G}+\mathbf{x_{NC}}$).
In Fig.~\ref{fig:meto_syg_spec} the spectrogram of the simulated signal $\mathbf{x}$ and the spectrogram of the additional simulated realization of the signal $\mathbf{x}$ called $\mathbf{x'}$ (with the same parameters) are presented. It is made to emphasize that during simulation the main difference between the simulated signals (with the same parameters) will be noticeable in the amount, time stamps, and energy of non-cyclic impulses (see Fig. \ref{fig:meto_syg_spec} marked with black arrows). This behavior corresponds to the idea of signal segmentation. Dividing a signal with a local fault (to obtain shorter segments), we have the same information about the local fault in each segment; however, the level of interference caused by non-cyclic impulses is different. This is a crucial difference for the proposed methodology, as the subsignals of the spectrogram are tested for the appearance of the correlation. The higher the number of high-energy non-cyclic impulses, the more likely the carrier bands where they are spread will be highly correlated. Dividing the signal into shorter segments, the aim is to suppress this effect in the CM results. 

For the purpose of the paper, the 1000 MC simulations of the signal $\mathbf{x}$ are performed to obtain the averaged (statistically consistent) IFB selector for each measure of correlation. MC simulation is also repeated for various combinations of parameters of the simulated signal, i.e. the amplitudes of the non-cyclic impulses and cyclic impulses take values from $[10:5:30]$ and $[2:1:6]$. The ENVS indicator of CM-based IFB selectors is then calculated, as well as the computational costs of the consideblack CMs to score the usage of the proposed measures of the correlation, the median filtering, and the design of an averaged IFB selector. 

\section{Results}
\label{results}

The proposed methodology is examined for simulation data with different SNRs, by changing the SOI amplitude versus constant parameters of Gaussian noise and by changing the amplitude of non-cyclic impulses. The proposed approach is also tested for the real vibration data \JHSS{that come from hummer crusher working in the mining industry} and \JHSS{acoustic data of the bearing of the test rig electric motor}. The comparison with known methods is also visualized.

\subsection{\textcolor{black}{Simulated data analysis}}
\label{results_sim}

In this section, the proposed IFB selection method is tested for the $K=1000$ simulated 1 second signals, described in Section~\ref{model} (ACI = 3, ANCI = 20, $f_{soi}=30$ Hz, $cf_{nc}=6000$ Hz, $cf_{soi}=2500$ Hz). First, the exemplary result of the CM-based IFB selector design and data filtration is presented for a single simulated 1 second signal.  

\begin{figure}[ht!]
  \centering
\includegraphics[width=0.7\textwidth]{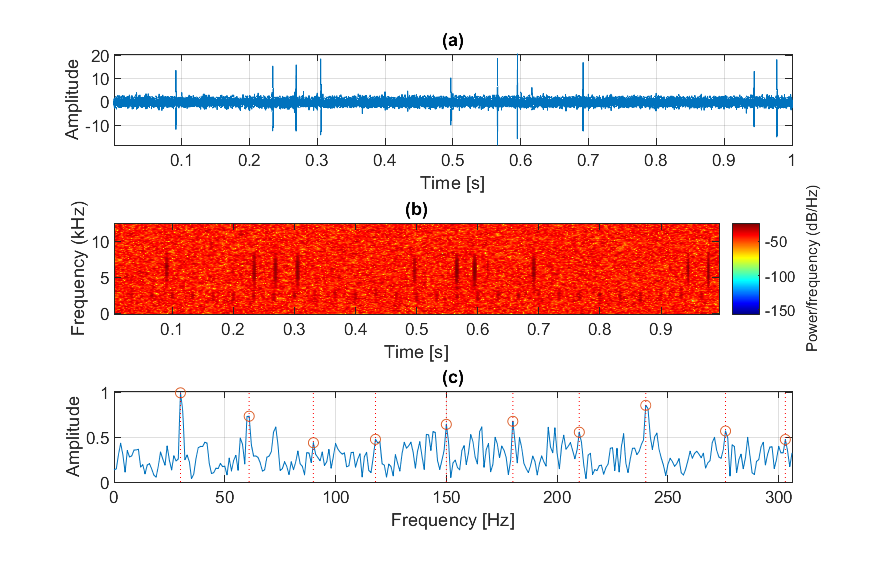}
    \caption{(a) Simulated signal, (b) its spectrogram and (c) SES with weak fault related harmonics marked in black circles.}
    \label{simulated}
\end{figure}

\begin{figure}[t!]
    \centering
    \begin{subfigure}[t]{0.5\textwidth}
        \centering
\includegraphics[width=0.99\textwidth]{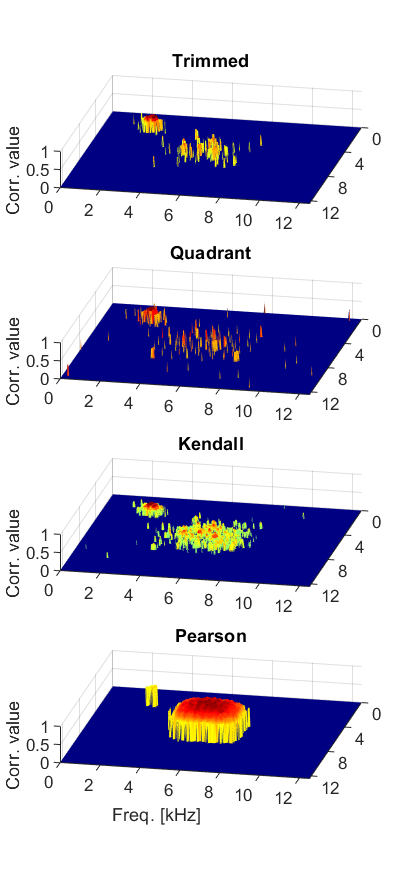}
        \caption{}
        \label{fig:maps_a}
    \end{subfigure}%
    ~ 
    \begin{subfigure}[t]{0.5\textwidth}
        \centering
\includegraphics[width=0.99\textwidth]{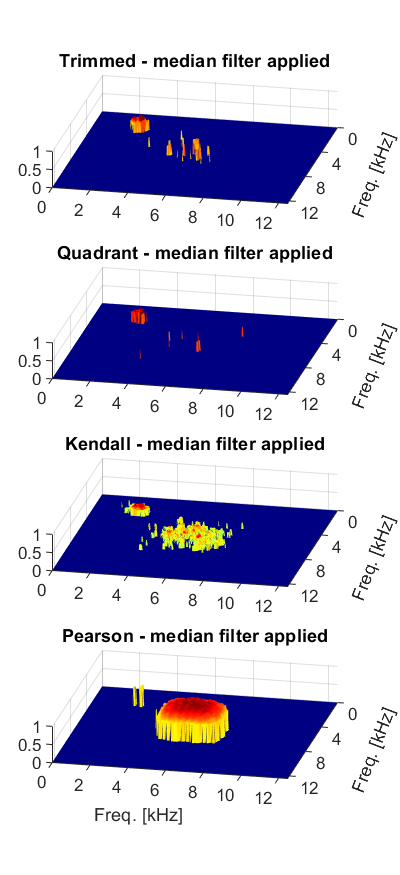}
        \caption{}
    \end{subfigure}
    \caption{Comparison of the results of the proposed correlation measures and median filter utilization i.e. (a) correlation maps (CMs) for simulated signal and (b) corresponding CMs after median filter application.}
    \label{fig:maps}
\end{figure}

The exemplary simulated 1 second signal and its spectrogram, see Eq.~(\ref{eq:stft}), as well as an SES are presented in Fig.~\ref{simulated}. 
The correlation measures, defined in Section~\ref{metodo}, were used to test the correlation of subsignals $subs(i)$ of the spectrogram corresponding to each frequency band $f_i$, see Eq.~(\ref{eq:sab}). Thus, trimmed, quadrant, Kendall, and Pearson CMs were obtained; see Fig.~\ref{fig:maps}.

As one can see in Fig.~\ref{fig:maps}, all maps pointed to the correct IFB around $2-3$ kHz with some disturbances in higher frequency ranges. One may notice that the higher value of the correlation presents the Pearson CM, that is, approximately $0.81$, see Fig.~\ref{fig:maps} bottom row. However, it is for the frequency band range around $5-7$ kHz, where the non-cyclic component of the impulsive noise appears. The misleading result is the consequence of the sensitivity of PCC for the outliers. The trimmed, quadrant, and Kendall CMs reveal an efficiency similar to each other, with some difference in the maximal value of the detected correlation (KCC has the smallest value in the IFB, approximately $0.25$). Trimmed and quadrant CMs seem to have fewer high-value correlations in the non-informative frequency bands ($4-8$ kHZ). They appear mainly as single-bin spikes. Fig.~\ref{fig:maps}(b) presents the results of the median filter application. One can observe the smoothing of the maps. Elimination of unwanted high correlation values in the $4-8$ kHz range is especially visible for trimmed and quadrant CMs. 
It should be noted that only the PCC takes a value close to 1, equal to $0.81$, and other correlation measures take values much lower (approximately $0.25-0.4$). The TCC and QCC seem to be more resistant to outliers, but their response to cyclic information is weaker than that of the PCC. However, normalization (by maximum values), performed in the next step during the IFB selector design, flattens this difference.

\begin{figure}[ht!]
  \centering
\includegraphics[width=0.8\textwidth]{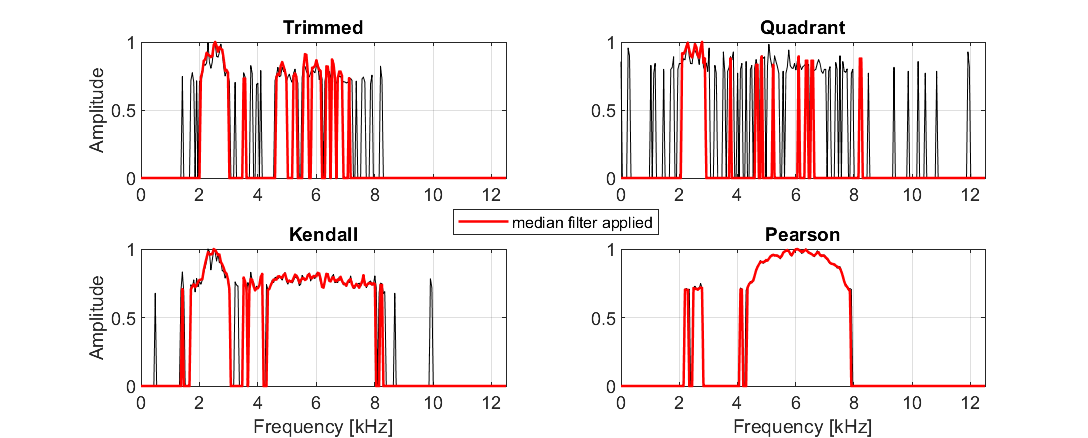}
    \caption{\textcolor{black}{Pearson, Kendall, quadrant, and trimmed CM-based IFB selectors with the corresponding results of IFB selectors for the CMs with the median filter applied for the simulated signal.}}
    \label{fig:sele}
\end{figure}

To simplify and compress the information contained in the 2D map of correlations in the next step, each column of CM is integrated into a single value, creating a 1D IFB selector. The results of the IFB selector, which plays the role of filter characteristic, are shown in Fig.~\ref{fig:sele}. 
As one can see, the Pearson IFB selector incorrectly pointed out the informative frequency band as the frequency band around 6 kHz, which corresponds to the existence of non-cyclic impulses.
Unlike the new IFB selectors (based on TCC and QCC), which have the highest values around 2.5 kHz, it correctly highlights the IFB, similar to the results of the Kendall IFB selector. The results of the IFB selectors based on KCC, QCC, and TCC differ especially in the range of $4-8$ kHz; there are some narrow bands with correlation values different from 0, see Fig.~\ref{fig:sele} black line, which is an unwanted phenomenon for an ideal selector. Therefore, the median filter is applied in CM, with the aim of smoothing out the results and removing some outlier values. The results of the IFB selector after the application of the median filter are marked with a black line in Fig.~\ref{fig:sele}. As mentioned, the median filter contributed to smoothing the results of the CM, and consequently improved the trimmed and quadrant IFB selectors. It removed a significant part of the values above zero in the non-informative frequency band ($4-8$ kHz). Kendall IFB selector result is also improved (see left bottom panel in Fig. \ref{fig:sele}), some high values in the non-informative part of the frequency range were removed. In the case of the Pearson IFB selector, there is almost no difference in the results of the IFB selectors before and after the median filter application. This is related to the high sensitivity of this measure to outliers of the signal; the median filter did not remove high amplitudes in the frequency bands around $4-8$ kHz, as there were too many of them and do not occur as single high values (no big difference in the neighborhood of 3-by-3 around the corresponding value in the CM).

\begin{figure}[ht!]
 \centering
\includegraphics[width=0.8\textwidth]{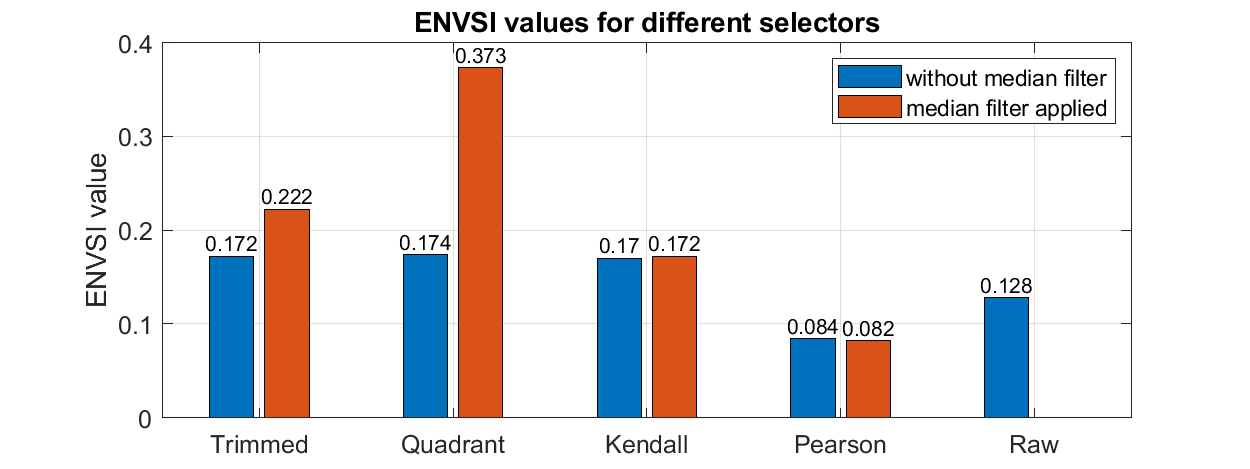}
    \caption{ENVSI values for different IFB selectors based on the CMs without median filter (blue bars) applied and with the median filter (black bars) applied for the simulated signal.}
    \label{tab1}
\end{figure}

\begin{figure}[ht!]
  \centering
\includegraphics[width=0.8\textwidth]{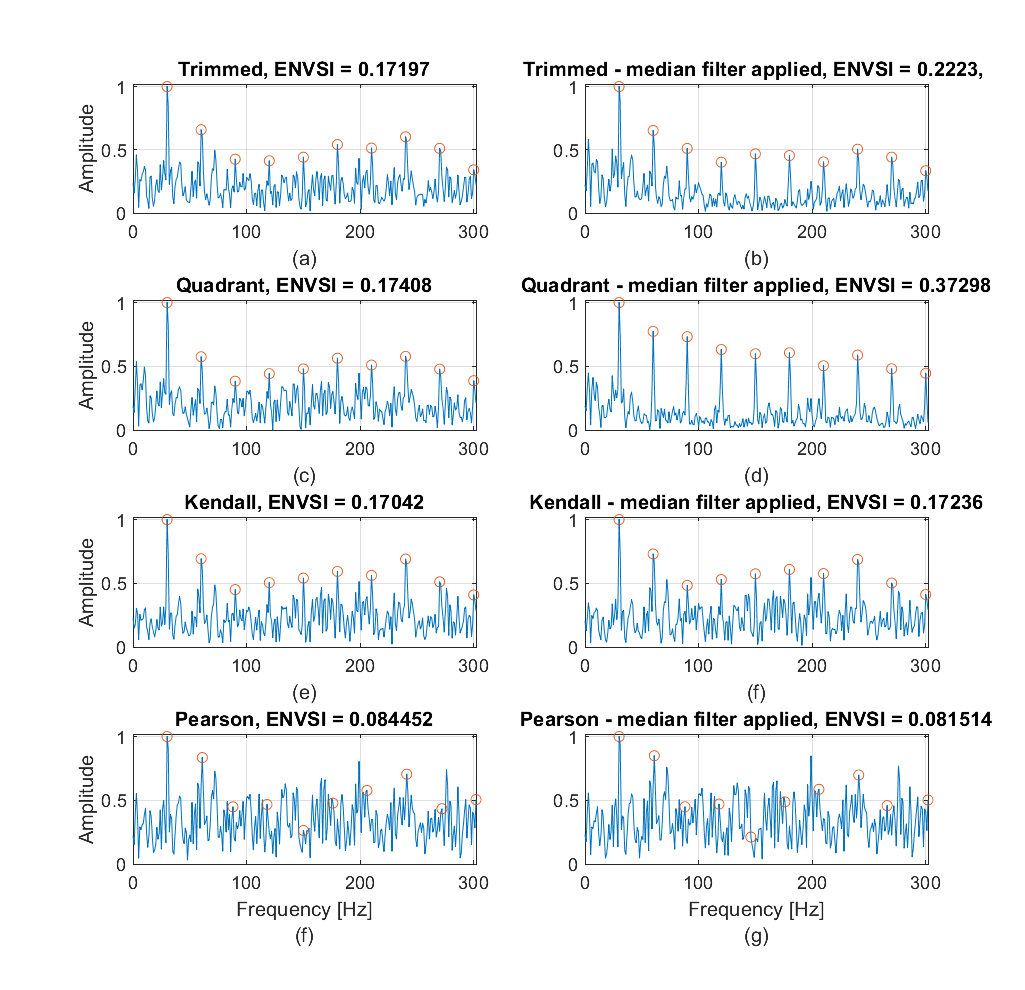}
    \caption{SES of the signal filteblack with the different IFB selectors: a) trimmed IFB selector b) trimmed IFB selector with the median filter applied c) quadrant IFB selector d) quadrant IFB selector with the median filter applied e) Kendall IFB selector f) Kendall IFB selector with the median filter applied g) Pearson IFB selector h) Pearson IFB selector with the median filter applied for the simulated signal.}
    \label{fig:envsi_sym1000}
\end{figure}

The IFB selectors presented in Fig.~\ref{fig:sele} were used to perform data filtration and to compare the results of the correlation measures and the median filter utilization used in CM using ENVSI. The ENVSI value for the raw simulated signal is equal to 0.1282, which is a portion of the fault harmonics (10 first harmonics included, according to \cite{Hebda-Sobkowicz2020}) in the squablack envelope spectrum (SES) distribution to the entire SES in the consideblack range; see Fig.~\ref{simulated}, bottom panel. The use of an effective IFB selector means that the ENVSI value will grow after filtration. 
The results of ENVSI for the trimmed, quadrant, Kendall and Pearson IFB selectors are presented in Fig.~\ref{tab1}, where the blue bars denote ENVSI values for the IFB selectors based on CM without the median filter applied during the CM calculation, and the black bars represent ENVSI values calculated for the IFB selectors based on CM with the median filter application. The corresponding SES are presented in Fig.~\ref{fig:envsi_sym1000}, respectively. 

The highest value of ENVSI, before the median filter application, equal to 0.1741 takes a trimmed IFB selector. The ENVSI value for the Pearson IFB selector (ENVSI $=0.0845$) is lower than that of the raw signal, which means that filtration with this IFB selector is insufficient and makes damage detection more difficult than it could be using the raw signal. As mentioned above, this is a consequence of the PCC sensitivity to high non-cyclic impulses in the data. Pearson IFB selector focuses on the non-informative part, related to impulsive noise. There is no cyclic component, and the envelope spectrum does not contain information about the fault and fault harmonics, so the ENVSI value of the filteblack signal with this IFB selector is reasonably lower than that of the raw signal. 

As can be seen in Fig.~\ref{tab1}, considering the CMs without the applied median filter, the effectiveness of the quadrant, trimmed, and Kendall IFB selectors is similar (with a slight pblackominance of the first two), but there is a significant improvement in ENVSI values compablack to the ENVSI values of the Pearson IFB selector and the raw signal. 
However, the median filter application in the case of REs of CM improved the results. The best improvement after the median filter application is made for the quadrant IFB selector, where the filtration efficiency has increased by approximately $114\%$ ($29\%$ in the case of the trimmed IFB selector) and compablack to the raw signal, filtration with the quadrant IFB selector improves the quality of the SES to $191\%$ ($73\%$ in the case of the trimmed IFB selector).

The methodology was repeated during 1000 MC simulations (1000 simulated signals, with the same parameters, described in Section~\ref{model}, with ACI $=3$ and ANCI $=20$) to obtain averaged results that present the global tendency of the presented procedure of the new application of REs of the correlation and median filter utilization in the CMs. As the final result of the MC simulation, the CM-based averaged IFB selector will be calculated from the 1000 CM-based IFB selectors. 

\begin{figure}[ht!]
 \centering
\includegraphics[width=0.85\textwidth]{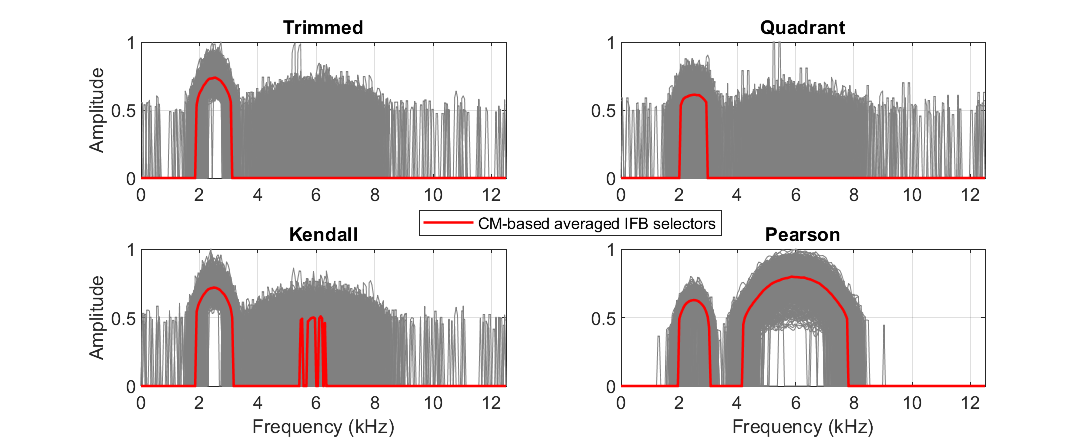}
    \caption{Pearson, Kendall, quadrant, and trimmed CM-based averaged IFB selectors for simulated signals.}
    \label{fig:sele1}
\end{figure}

Fig.~\ref{fig:sele1} presents the CM-based IFB selectors of the 1000 MC simulation (marked in gray lines, overlapping each other) and the CM-based averaged IFB selector (marked in the black line, the median was used as a more robust average measure).
As one can observe, the single IFB selector (derived from 1 second simulated signal) still can have some high values of correlation in the frequency bands not related to the IFB even when the median filter is applied.
Therefore, the averaged IFB selector (from 1000 IFB selectors based on the CMs with the median filter applied) tends to smooth the results and remove high values of the correlation related to non-informative impulsive subsignals. In the case of the trimmed, quadrant and Kendall IFB selectors, the averaged (in the median sense) IFB selectors improve their results by improving selectivity around IFB and removing (trimmed and quadrant IFB selectors) or blackucing (Pearson, Kendall IFB selectors) high values of IFB selectors in non-informative frequency bands, between $4-8$ kHz where the non-cyclic impulses occur.

\begin{figure}[ht!]
 \centering
\includegraphics[width=0.85\textwidth]{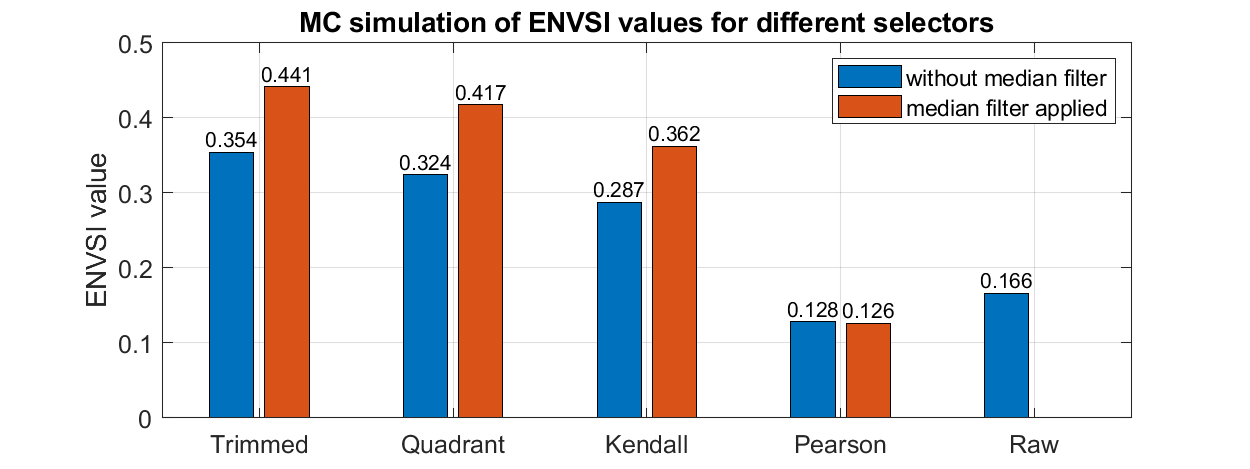}
    \caption{ENVSI values calculated for 1000 MC simulation of different CM-based with median filter applied IFB selectors.}
    \label{tab2}
\end{figure}

In Fig.~\ref{tab2}, the application of the median filter in CMs is compablack to the results of CMs without the application of the median filter for 1000 MC simulations. For IFB selectors with REs of the correlation, the median filter improves the result of the signal filtration, and the ENVSI value increased. Despite the Pearson IFB selector, where no improvement is observed, as was also discussed in the case of a single-signal simulation $\mathbf{x}$. 

 MC simulation shows that the most efficient IFB selector is a trimmed IFB selector, and the median filter application in CM even improves its results.
As can be seen in Fig.~\ref{tab2}, the use of TCC and the median filter application in CM allows one to increase the ENVSI value (which corresponds to an improvement in the quality of the SES of the raw signal) by approximately $196\%$ compablack to the raw signal. Similar effectiveness occurs for the quadrant IFB selector.  The MC simulation confirms that the Pearson IFB selector is not dedicated to signals with non-Gaussian noise, and its usage makes detection even more difficult (the median value of ENVSI for 1000 signals is approximately $23\%$ lower than it is for the raw signal).

\begin{figure*}[ht!]
\centering
\begin{subfigure}[b]{0.48\textwidth}
\includegraphics[width=1\textwidth]{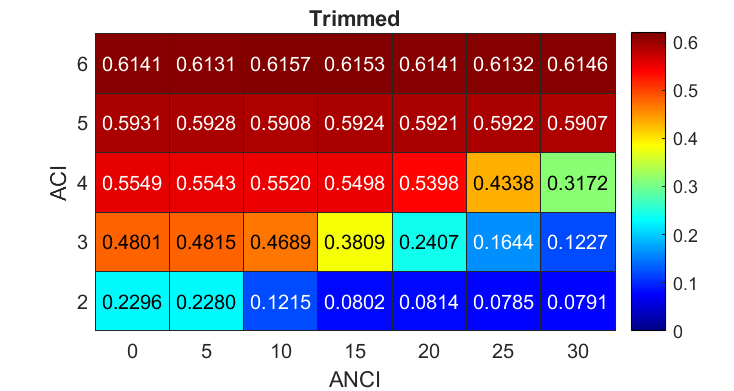}
\caption{} 
\end{subfigure} \quad
\begin{subfigure}[b]{0.48\textwidth}
\includegraphics[width=1\textwidth]{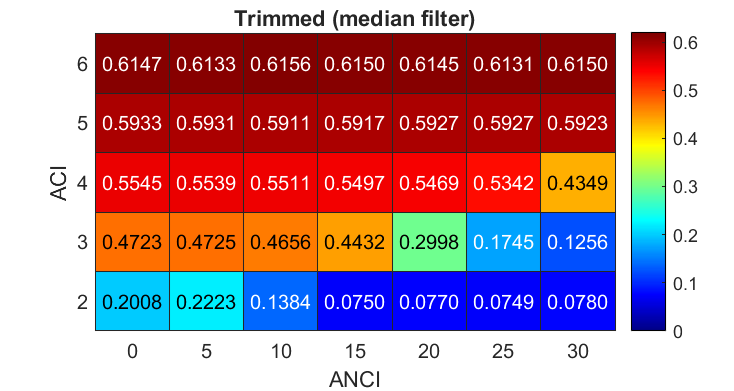}
\caption{}
\end{subfigure} 
\\
\begin{subfigure}[b]{0.48\textwidth}
\includegraphics[width=1\textwidth]{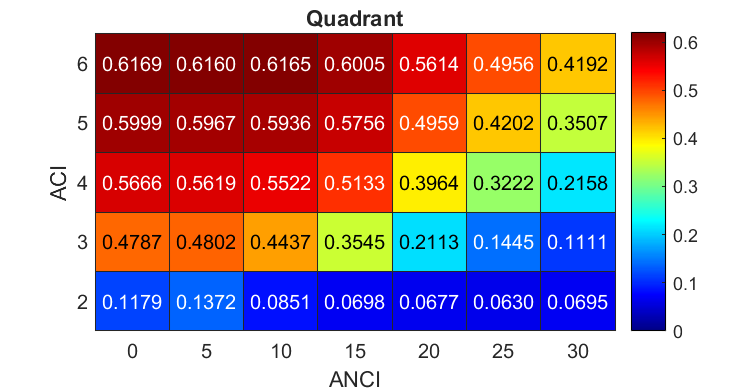}
\caption{}
\end{subfigure}\quad
\begin{subfigure}[b]{0.48\textwidth}
\includegraphics[width=1\textwidth]{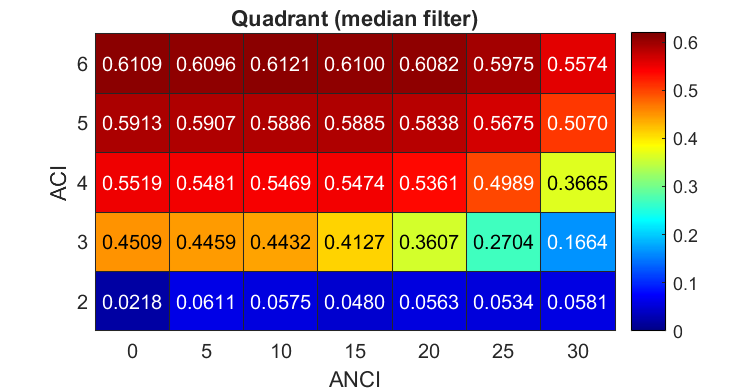}
\caption{}
\end{subfigure}
\\
\begin{subfigure}[b]{0.48\textwidth}
\includegraphics[width=1\textwidth]{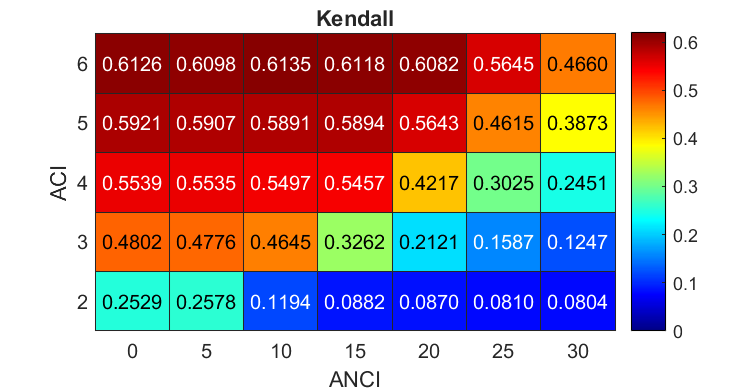}
\caption{}
\end{subfigure}\quad
\begin{subfigure}[b]{0.48\textwidth}
\includegraphics[width=1\textwidth]{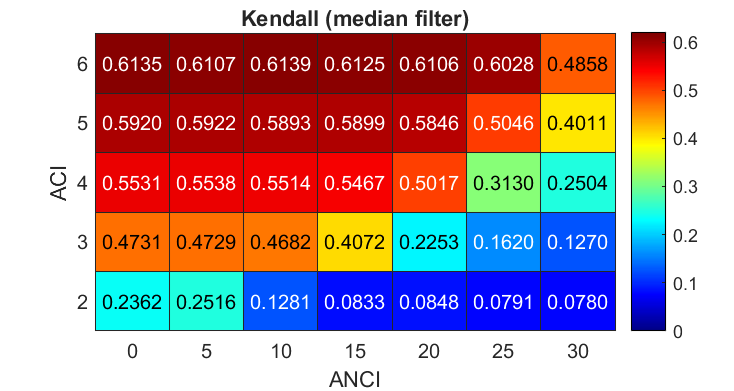}
\caption{}
\end{subfigure}
\\
\begin{subfigure}[b]{0.48\textwidth}
\includegraphics[width=1\textwidth]{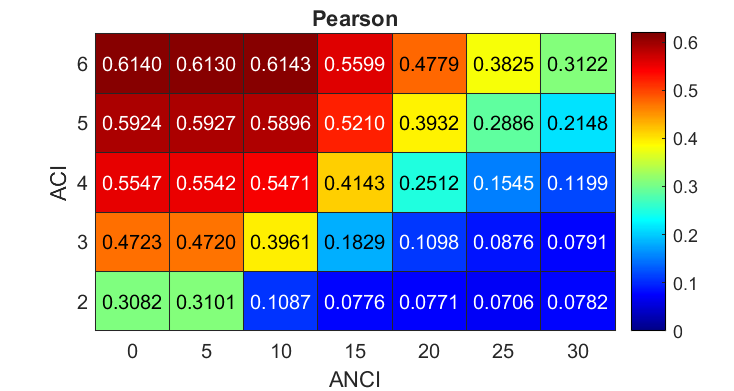}
\caption{}
\end{subfigure}\quad
\begin{subfigure}[b]{0.48\textwidth}
\includegraphics[width=1\textwidth]{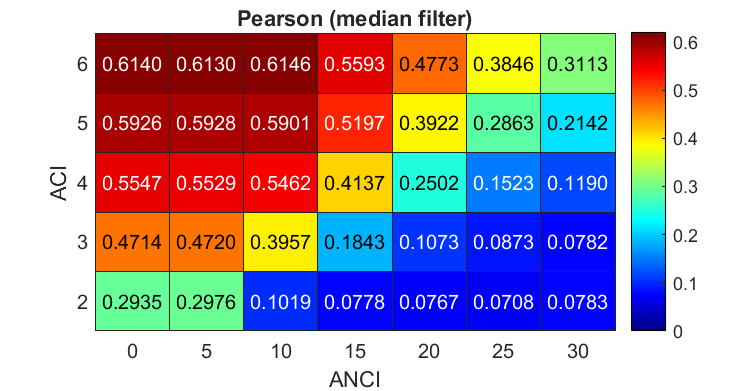}
\caption{}
\end{subfigure}
\caption{\textcolor{black}{ENVSI values calculated for 1000 MC simulation of different IFB selectors: a) trimmed IFB selector b) trimmed IFB selector with the median filter applied c) quadrant IFB selector d) quadrant IFB selector with the median filter applied e) Kendall IFB selector f) Kendall IFB selector with the median filter applied g) Pearson IFB selector h) Pearson IFB selector with the median filter applied and changing parameters of simulated signals.}}
\label{fig:map_sim}
\end{figure*}

The effectiveness of the correlation measures, as well as the median filter utilization in the CM, \JHSS{was} also checked for various combinations of parameters in the simulated signal. During the MC simulation, the amplitudes of the non-cyclic impulses and cyclic impulses take values from $[10:5:30]$ and $[2:1:6]$, respectively. Hence, the ratio of cyclic to non-cyclic impulses changes from $\frac{1}{5}$ to $\frac{1}{15}$ (in the worst case, the non-cyclic impulses are 15 times higher than the cyclic impulses). The parameters of the background Gaussian white noise are constant. The results of the ENVSI values of the 1000 MC simulation for different IFB selectors and for changing parameters of the simulated signal $\mathbf{x}$ are presented in Fig.~\ref{fig:map_sim}. 

\begin{figure}[ht!]
  \centering
\includegraphics[width=0.7\textwidth]{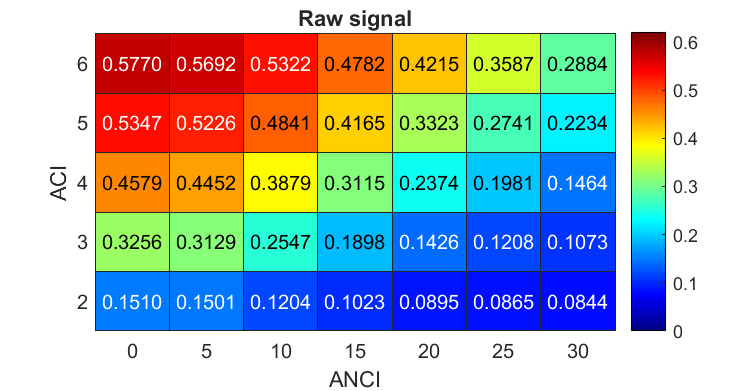}
    \caption{\textcolor{black}{ENVSI values calculated for 1000 MC simulation of raw signals.}}
    \label{fig:map_sim1}
\end{figure}

The average value of ENVSI from the MC simulation of the raw signal is presented in Fig.~\ref{fig:map_sim1}. 
As can be seen for ACI = 2 (cyclic impulses completely hidden in Gaussian white noise) and ANCI $\geq$ 15 the tested correlation measures do not detect (ENVSI value $\leq$ 0.1) the periodicity associated with the local fault. In the case of Gaussian white noise and ACI = 2 (without non-cyclic impulses), the best effectiveness returns the Pearson IFB selector, then Kendall and trimmed IFB selectors. The quadrant IFB selector seems to be insufficient in this case (the ENVSI value is smaller than its value for the raw signal, see Fig.~\ref{fig:map_sim1}). The results improve for the cyclic component with an amplitude of 3 ($\sigma=1$, so the values of the noise amplitudes vary between $[-3,3]$, and the impulses can still be hidden in Gaussian white noise). As one can see, in Fig.~\ref{fig:map_sim} subplots on the right i.e. subplots Fig.~\ref{fig:map_sim}(b), Fig.~\ref{fig:map_sim}(d), Fig.~\ref{fig:map_sim}(f), and Fig.~\ref{fig:map_sim}(h), the median filter can significantly improve the effectiveness of IFB selectors. It is especially noticeable in the case of the trimmed and quadrant IFB selectors, where it managed to increase its effectiveness by approximately $70\%$ (see Fig.~\ref{fig:map_sim}, plots (c) and (d), case ACI = $4$ and ANCI = $30$). In the case of the Pearson IFB selector, the median filter does not improve the results, as the Pearson IFB selector mostly focuses on the non-informative part of the frequency band where high-amplitude non-cyclic impulses appear. As one can see in Fig.~\ref{fig:map_sim} and Fig.~\ref{fig:map_sim1} the Pearson IFB selector improves the SES of the raw signal (higher ENVSI values than it does for the raw signal) for some cases of the signal parameters.
If the amplitude of non-cyclic impulses grows, then the Pearson IFB selector becomes insufficient and gives misleading results (creates a selector that, while filtering, selects mainly uninformative impulsive noise, which distorts the SES).

\begin{figure*}[ht!]
\centering
\begin{subfigure}[b]{0.45\textwidth}
\includegraphics[width=0.99\textwidth]{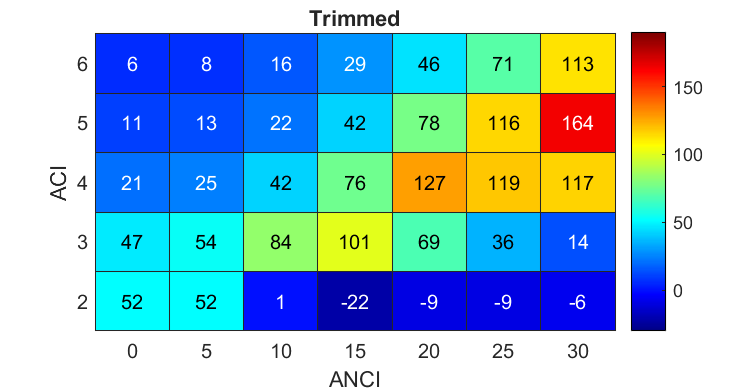}
\caption{} 
\end{subfigure} \quad
\begin{subfigure}[b]{0.45\textwidth}
\includegraphics[width=0.99\textwidth]{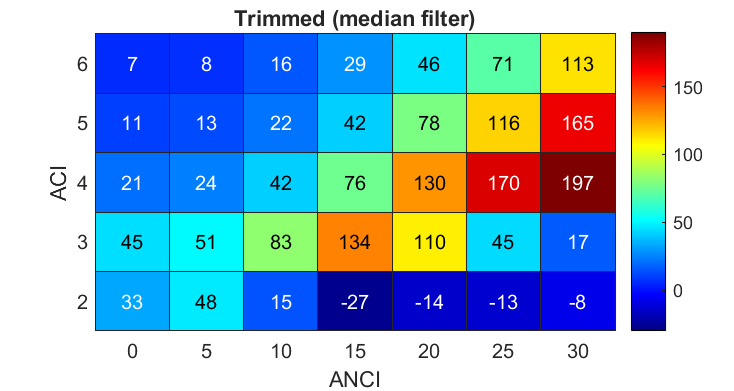}
\caption{}
\end{subfigure} 
\\
\begin{subfigure}[b]{0.45\textwidth}
\includegraphics[width=0.99\textwidth]{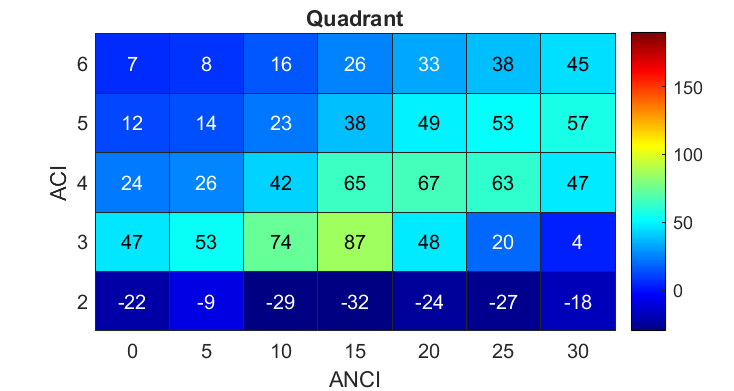}
\caption{}
\end{subfigure}\quad
\begin{subfigure}[b]{0.45\textwidth}
\includegraphics[width=0.99\textwidth]{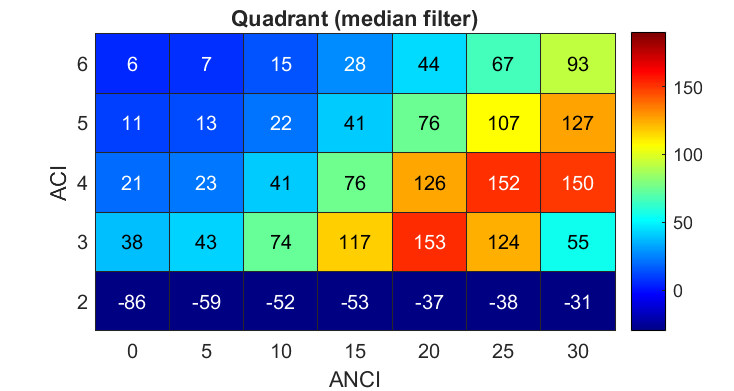}
\caption{}
\end{subfigure}
\\
\begin{subfigure}[b]{0.45\textwidth}
\includegraphics[width=0.99\textwidth]{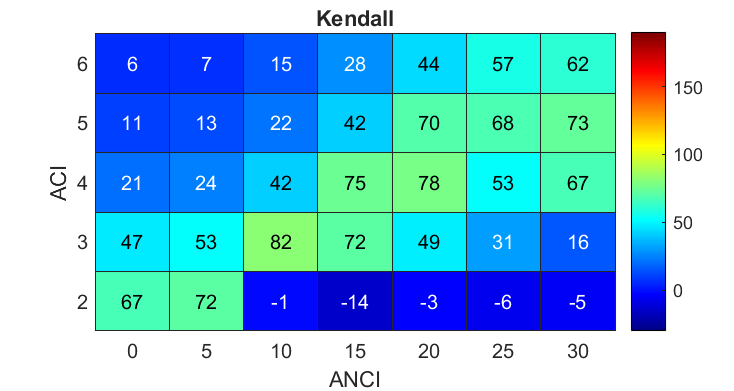}
\caption{}
\end{subfigure}\quad
\begin{subfigure}[b]{0.45\textwidth}
\includegraphics[width=0.99\textwidth]{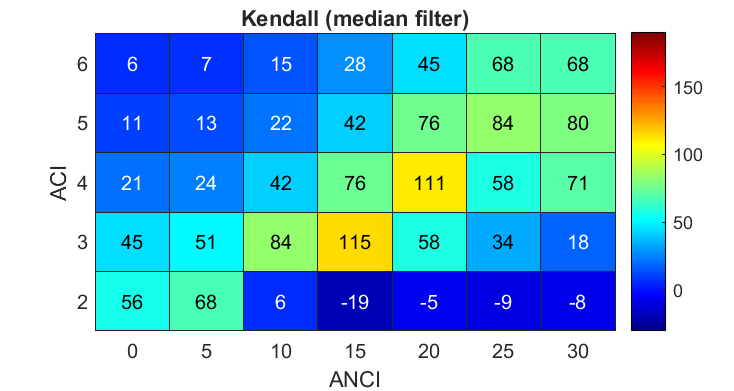}
\caption{}
\end{subfigure}
\\
\begin{subfigure}[b]{0.45\textwidth}
\includegraphics[width=0.99\textwidth]{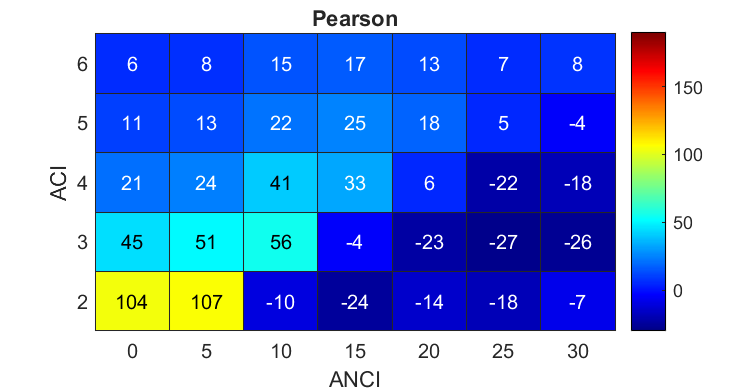}
\caption{}
\end{subfigure}\quad
\begin{subfigure}[b]{0.45\textwidth}
\includegraphics[width=0.99\textwidth]{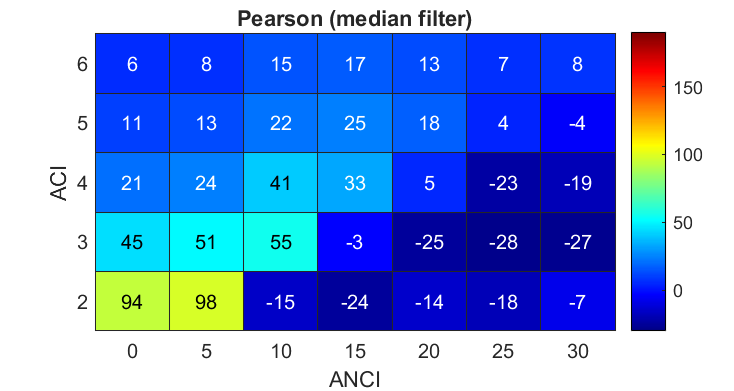}
\caption{}
\end{subfigure}
\caption{\textcolor{black}{ENVSI$_\text{score}$ improvement as a percentage value for different IFB selectors: a) trimmed IFB selector b) trimmed IFB selector with the median filter applied c) quadrant IFB selector d) quadrant IFB selector with the median filter applied e) Kendall IFB selector f) Kendall IFB selector with the median filter applied g) Pearson IFB selector h) Pearson IFB selector with the median filter applied and changing parameters of simulated signals.}}
\label{fig:map_sim2}
\end{figure*}

 For a better understanding of the results of the MC simulation, it is worth looking at the relationship between the difference of the ENVSI value of the filteblack and unfilteblack (raw) signal and the ENVSI value of the raw signal as the percentage value of quality improvement, i.e., $$\text{ENVSI}_{\text{score}}=\frac{\text{ENVSI(filteblack $s$)}-\text{ENVSI(raw $s$)}}{\text{ENVSI(raw $s$)}}*100\%,$$ where $s$ means signal. If the ratio is negative, it means that there is no improvement. As can be seen in Fig.\ref{fig:map_sim2} the median filter can significantly improve the quality of filtration, especially noticeable in the case of the quadrant IFB selector (see Fig.~\ref{fig:map_sim2}(c) and Fig.~\ref{fig:map_sim2}(d)).

\begin{figure}[ht!]
 \centering
\includegraphics[width=0.6\textwidth]{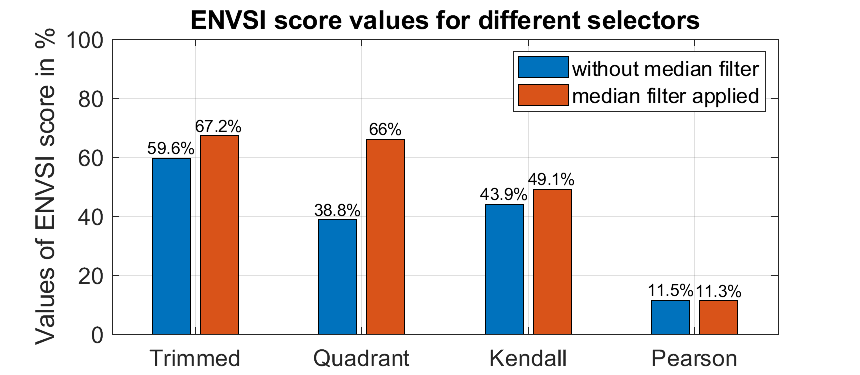}
    \caption{Average value of ENVSI$_\text{score}$ from MC simulation (including different values of ANCI and ACI $>2$) for different IFB selectors.}
    \label{tab2a}
\end{figure}

\textcolor{black}{Considering only ACI greater than 2  (all cases of the parameters of the simulated signals included, despite ACI $=2$) the averaged value of ENVSI for all IFB selectors is presented in Fig.~\ref{tab2a}. As one can see, the best performance (the average improvement of the ENVSI value up to $67.2\%$ - compablack to the ENVSI value of the raw signal) shows a trimmed IFB selector with the median filter applied during the CM calculation. In the case of the quadrant IFB selector, there is a high improvement in the ENVSI$_\text{score}$ value after applying the median filter (the results are almost twice as good if the median filter is used). In the case of the Kendall IFB selector, there is a small improvement in the ENVSI score value after the median filter utilization of about $5\%$. In the case of the Pearson IFB selector, there is no improvement in the ENVSI$_\text{score}$ results after the median filter application. }

\begin{figure*}[ht!]
\centering
\includegraphics[width=0.8\textwidth]{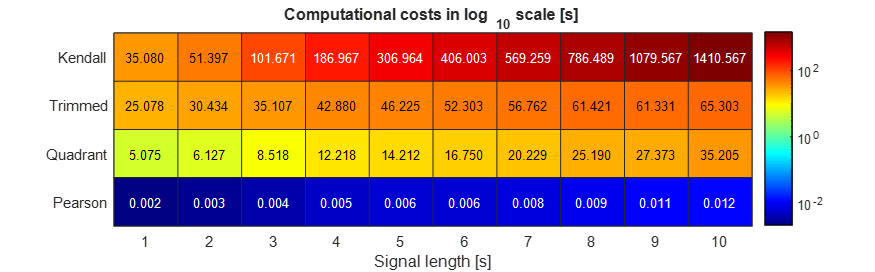}
\caption{The computational costs of 1000 MC simulation of the trimmed, quadrant, Kendall, and Pearson CMs for a different length of signal. } 
\label{fig:time1}
\end{figure*}

During 1000 MC simulations, the computational costs of the CMs calculation were also measublack for each measure of the correlation. The results considering changing the length of a signal are presented in Fig.~\ref{fig:time1}. 
It is worth notice that the CM matrix is symmetric along the diagonal, and only a calculation of half of the CM matrix is needed which enhances the calculation efficiency. In consequence, the operating time of all methods will decrease by half.
For a signal of 1 second length, the lowest computational cost occurs, as expected, for Pearson CM. One can conclude that the other REs of the correlation are modifications of the PCC; therefore, their computational costs will certainly be higher. However, the effectiveness of the Pearson CM for heavy-tailed signals is low. The CM that is effective during IFB selection for signals with non-Gaussian noise and relatively fast is the quadrant CM. Kendall CM has the highest computational cost. For a 10 second length signal, the computational cost of Kendall CM calculation is equal to 1410 seconds (23 minutes). It seems that the lower computational cost will occur for calculating 10 times the Kendall CM for a signal that lasts 1 second, than calculating it once for a 10-second signal.

\newpage
\subsection{\JHSS{Experiment description case 1 - vibration data of crusher machine}}
\label{sec:exp}
\begin{figure}[hbt!]
\centering
\includegraphics[width=0.45\textwidth]{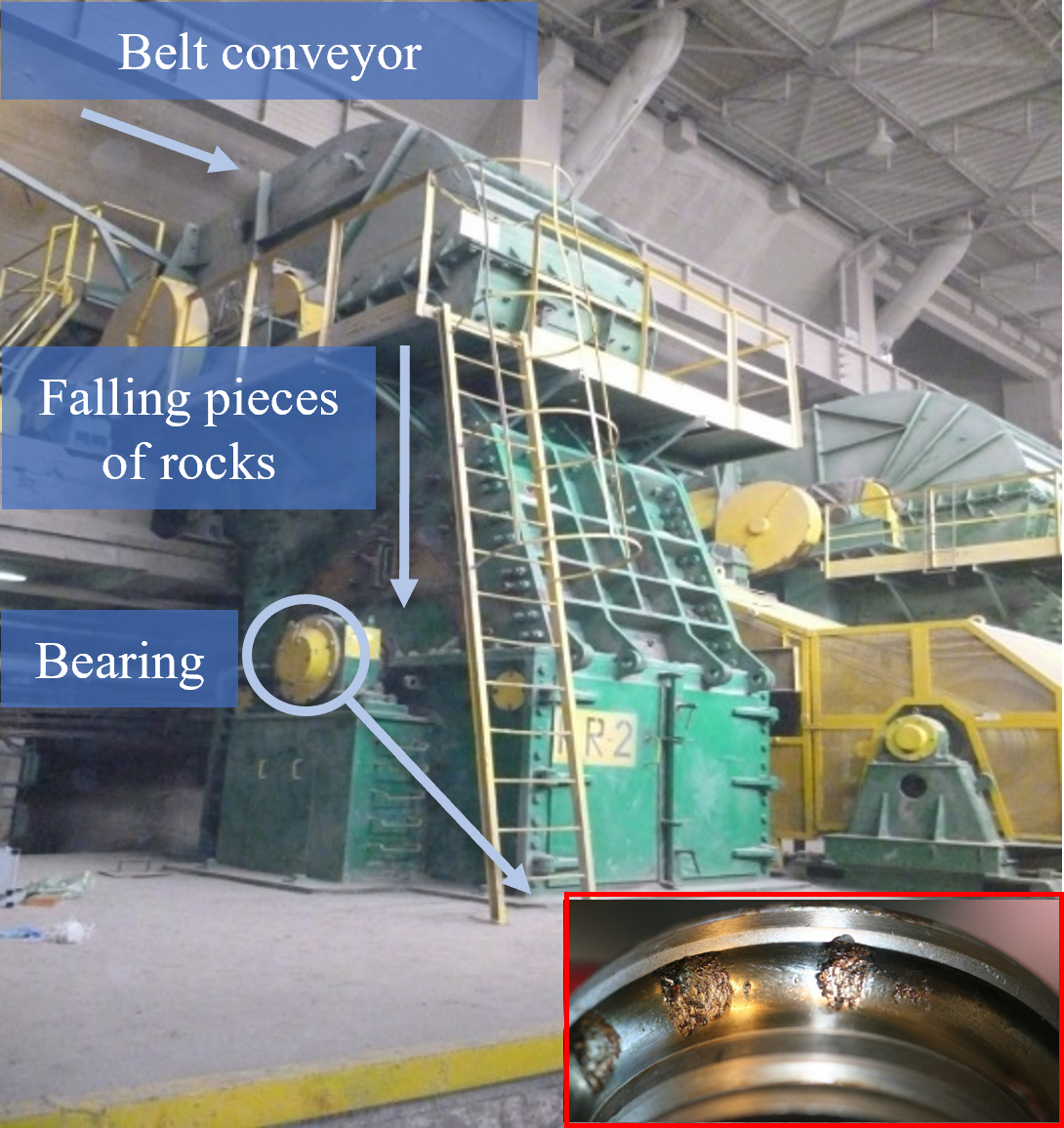}
\caption{Copper ore crusher \cite{wylomanska2016impulsive}}
\label{fig:obj_crush}
\end{figure}

The signal examined in this study is a vibration time series obtained from an ore crusher during standard operation in the mining industry (see Fig.~\ref{fig:obj_crush}).
 Data were collected at a sampling frequency of 25 kHz using Endevco accelerometers and a basic data acquisition set-up, which included an NI DAQ card and Labview Signal Express software. Throughout the recorded segments, the rotational speed can be assumed to remain relatively constant. The characteristic frequencies of the bearings are listed in Table \ref{tab:freqs}. The bearings under consideration are 23264 SKF. One of them exhibits localized damage to the inner race.

\begin{table}[ht!]
  \centering
      \caption{Characteristic frequencies of 23264 CCK/W33 bearing}
  	\resizebox{0.67\textwidth}{!}{
  \begin{tabular}{|l|l|}
  \hline
     \textbf{Description} & \textbf{Value} \\ \hline
     Rotational frequency of the inner ring & 3 Hz\\ \hline
     Rotational freq. of the rolling element and cage assembly & 1.3 Hz\\ \hline
     Rotational freq. of a rolling element about its own axis & 10.6 Hz\\ \hline
     \textbf{Over-rolling frequency of one point on the inner ring} & \textbf{30.7 Hz}\\ \hline
     Over-rolling frequency of one point on the outer ring & 23.3 Hz\\ \hline
     Over-rolling frequency of one point on a rolling element & 21.1 Hz\\ \hline
  \end{tabular}
  }
  \label{tab:freqs}
\end{table}

\subsection{\JHSS{Real vibration data analysis - bearing of the crusher machine}}
\label{sec:real}
\begin{figure}[ht!]
  \centering
    \includegraphics[width=0.7\textwidth]{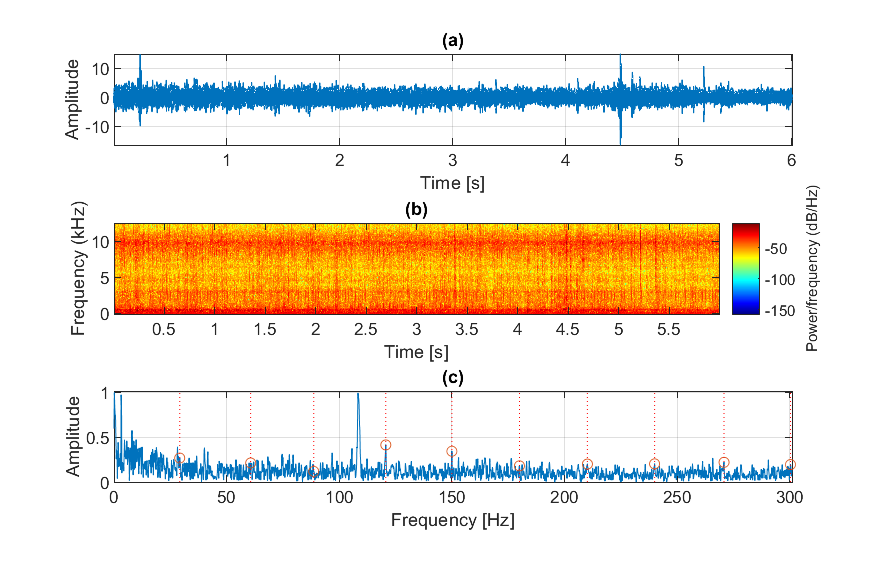}
    \caption{(a) Real vibration signal, (b) its spectrogram and (c) SES with weak fault related harmonics marked in black circles.}
    \label{fig:real}
\end{figure}

 \color{black} 
The real vibration signal, presented in Fig.~\ref{fig:real}, originates from the bearing of the crushing machine (hammer crusher), described in Section \ref{sec:exp}. 
Hammer crusher technology is characterized by an impulsive (non-Gaussian) noise, as a consequence of falling pieces of copper ore, then the non-cyclic impulses appear in the vibration signal that are not related to a damage. This makes the diagnostic procedure more difficult. The signal has a duration of 6 seconds with a sampling frequency equal to 25 kHz. The fault frequency is 30 Hz with a central carrier frequency around 2.5 kHz. 

The vibration signal and its spectrogram and SES are visible in Fig.~\ref{fig:real}. 
Impulses related to the fault are not visible in the time domain nor in the time-frequency domain (i.e. spectrogram). The cyclic component with fault frequency (30 Hz) and its harmonics are also invisible in the SES (marked as black circles in Fig.~\ref{fig:real}, bottom plot). The ENVSI value of the raw signal is close to 0, equal to 0.0631, which means that there is no evidence in SES of the existence of a local fault, i.e., cyclic impulsive component in the analyzed signal.

 The main goal of the proposed methodology is performing filtration (the design of the IFB selector as a filter characteristic is crucial) to improve the SNR and make the SES clearer (improve peak visibility at 30 Hz and its harmonics at 60 Hz, 90 Hz, etc., corresponding to local fault frequency). 
In further steps, the effectiveness of the new proposed IFB selectors based on robust correlation measures is tested. 

According to the methodology, see Fig.~\ref{fig:block_diag}, the 6-second signal was split into $K=12$ segments (a half-second was assumed as the length of each segment of the signal) for the CM-based averaged IFB selector design. The division into 12 segments was chosen as a compromise between the length of the segment (to achieve the best possible resolution of the signals) and the possibly largest number of segments. The number of segments is crucial in the CM-based averaged IFB selector construction to suppress the non-informative components of the filter characteristic related to randomly occurring high-energy impulses.

\begin{figure}[t!]
    \centering
    \begin{subfigure}[t]{0.5\textwidth}
        \centering
\includegraphics[width=0.99\textwidth]{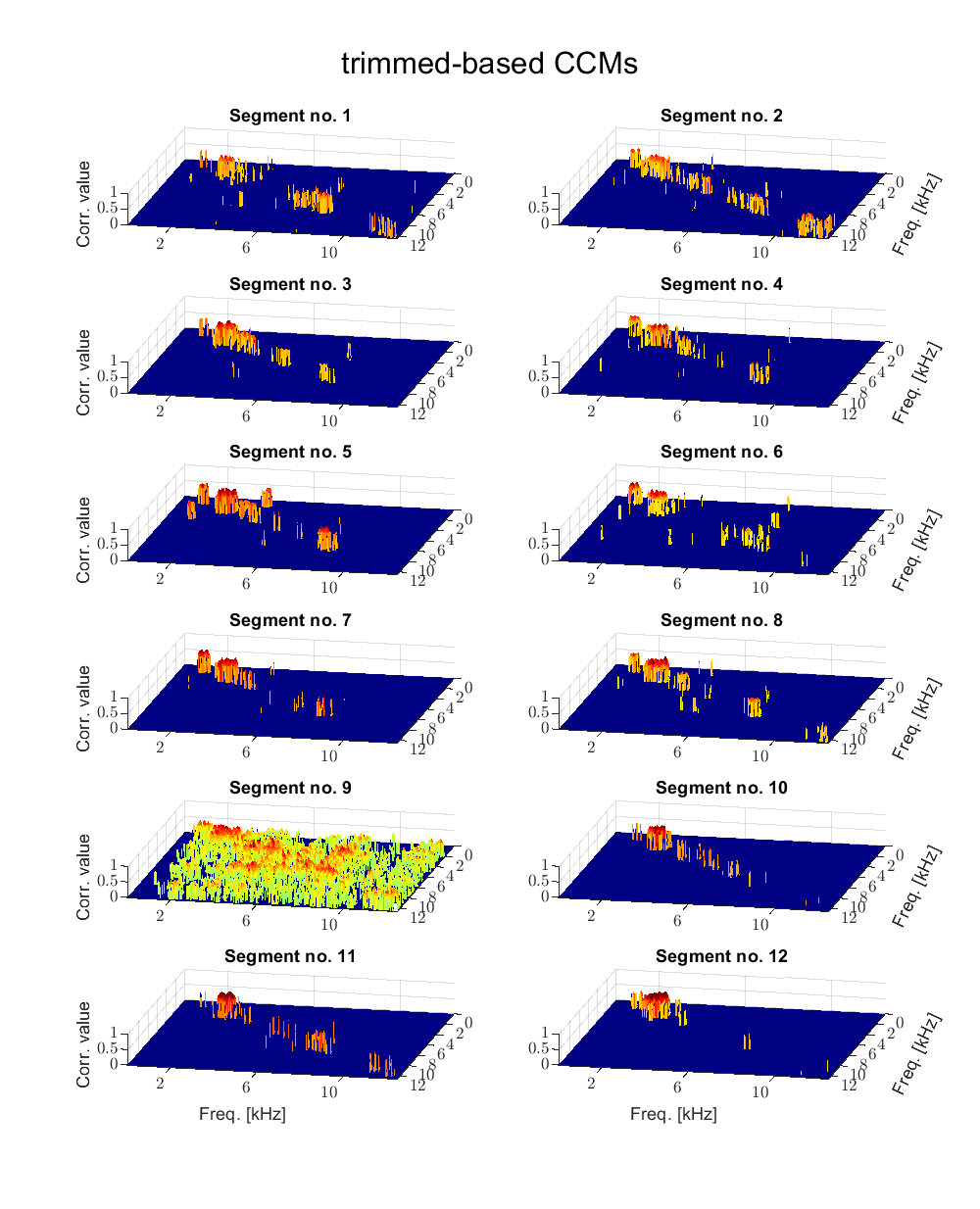}
        \caption{}
    \end{subfigure}%
    ~ 
    \begin{subfigure}[t]{0.5\textwidth}
        \centering
\includegraphics[width=0.99\textwidth]{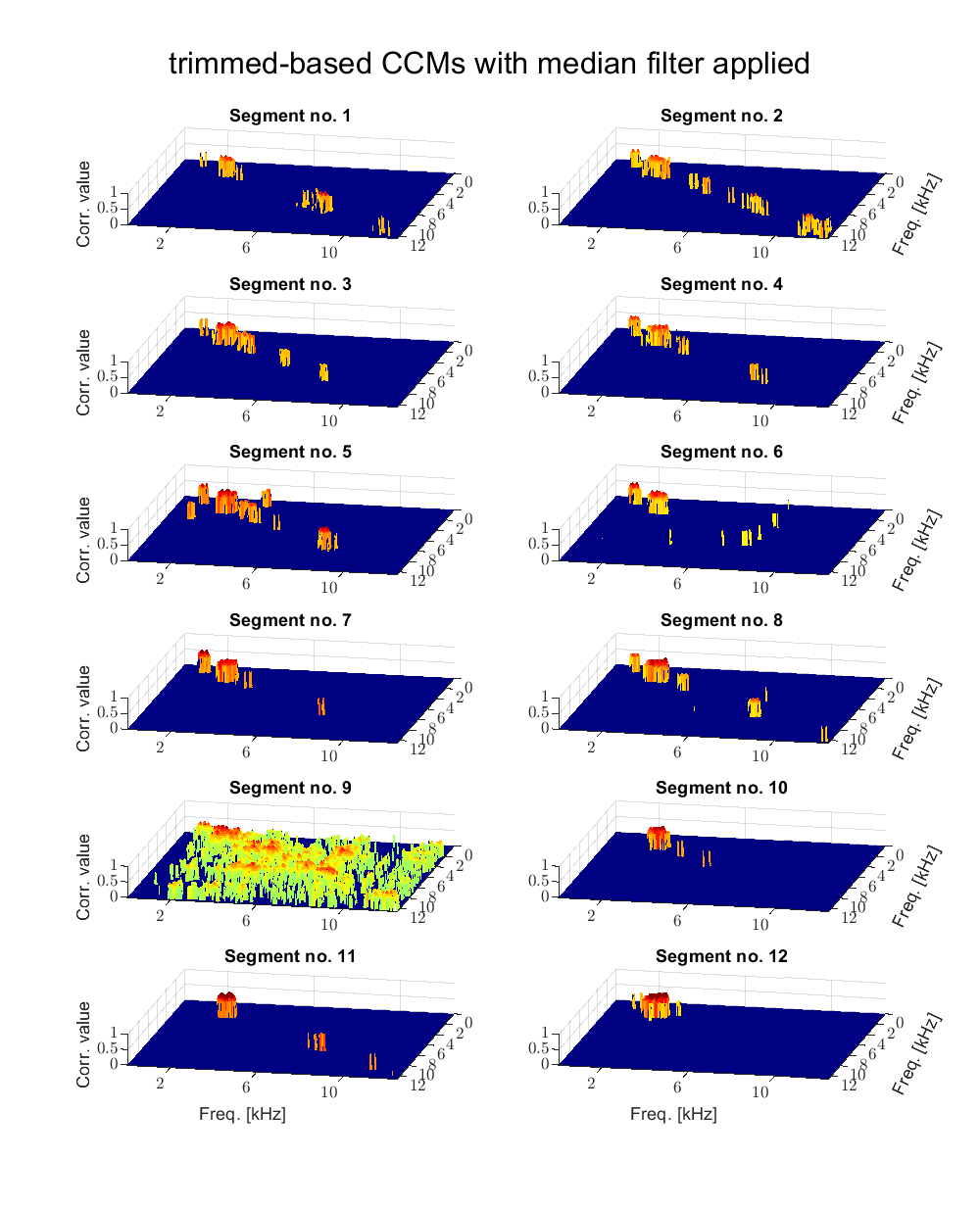}
        \caption{}
    \end{subfigure}
    \caption{Comparison of the results of the proposed correlation measures and median filter utilization i.e. (a) trimmed CMs for each segments of the real vibration signal and (b) corresponding CMs after median filter application.}
    \label{fig:real_maps_trim}
\end{figure}

\begin{figure}[t!]
    \centering
    \begin{subfigure}[t]{0.5\textwidth}
        \centering
\includegraphics[width=0.99\textwidth]{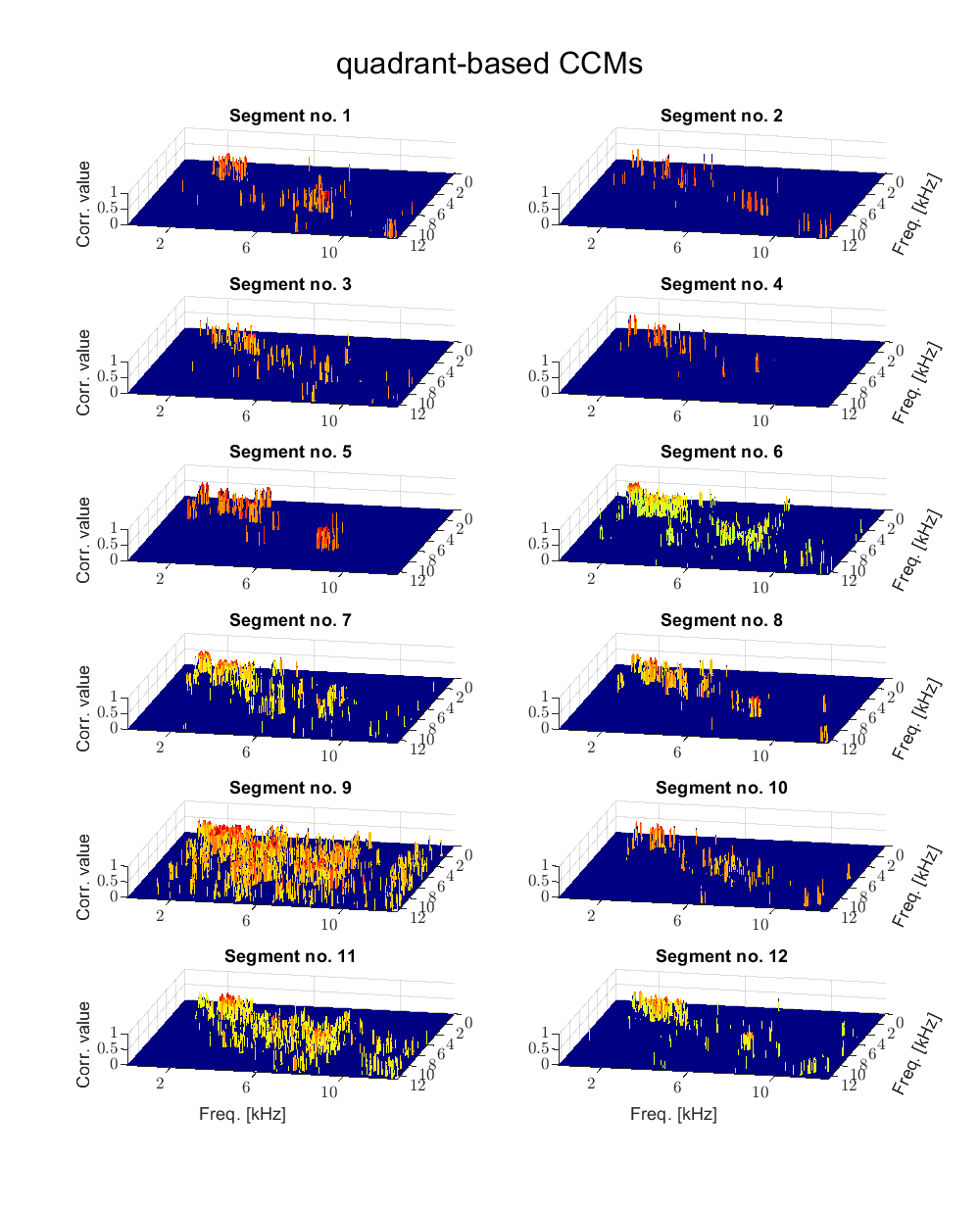}
        \caption{}
    \end{subfigure}%
    ~ 
    \begin{subfigure}[t]{0.5\textwidth}
        \centering
\includegraphics[width=0.99\textwidth]{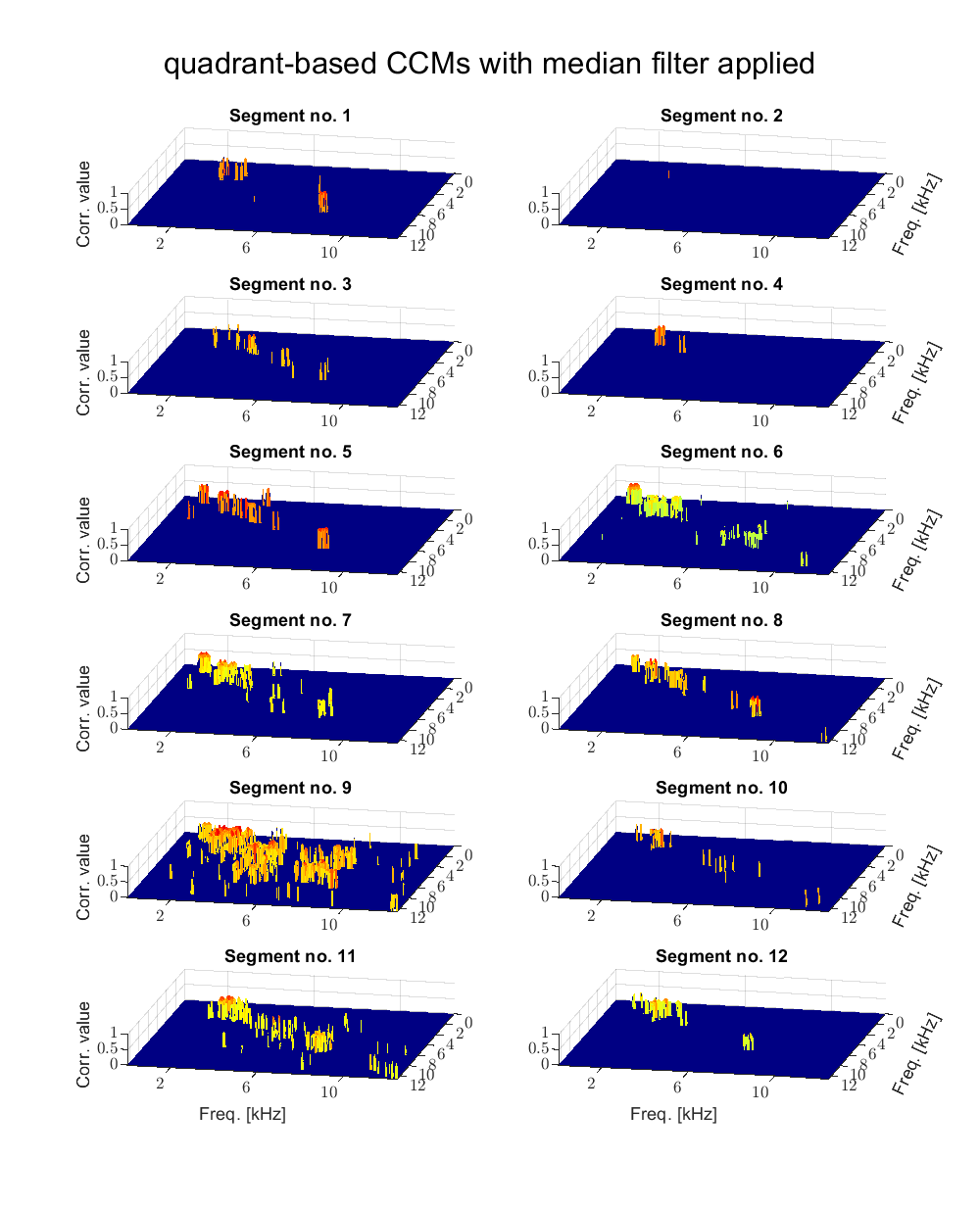}
        \caption{}
    \end{subfigure}
    \caption{Comparison of the results of the proposed correlation measures and median filter utilization i.e. (a) quadrant CMs for each segments of the real vibration signal and (b) corresponding CMs after median filter application.}
    \label{fig:real_maps_quad}
\end{figure}

\begin{figure}[t!]
    \centering
    \begin{subfigure}[t]{0.5\textwidth}
        \centering
\includegraphics[width=0.99\textwidth]{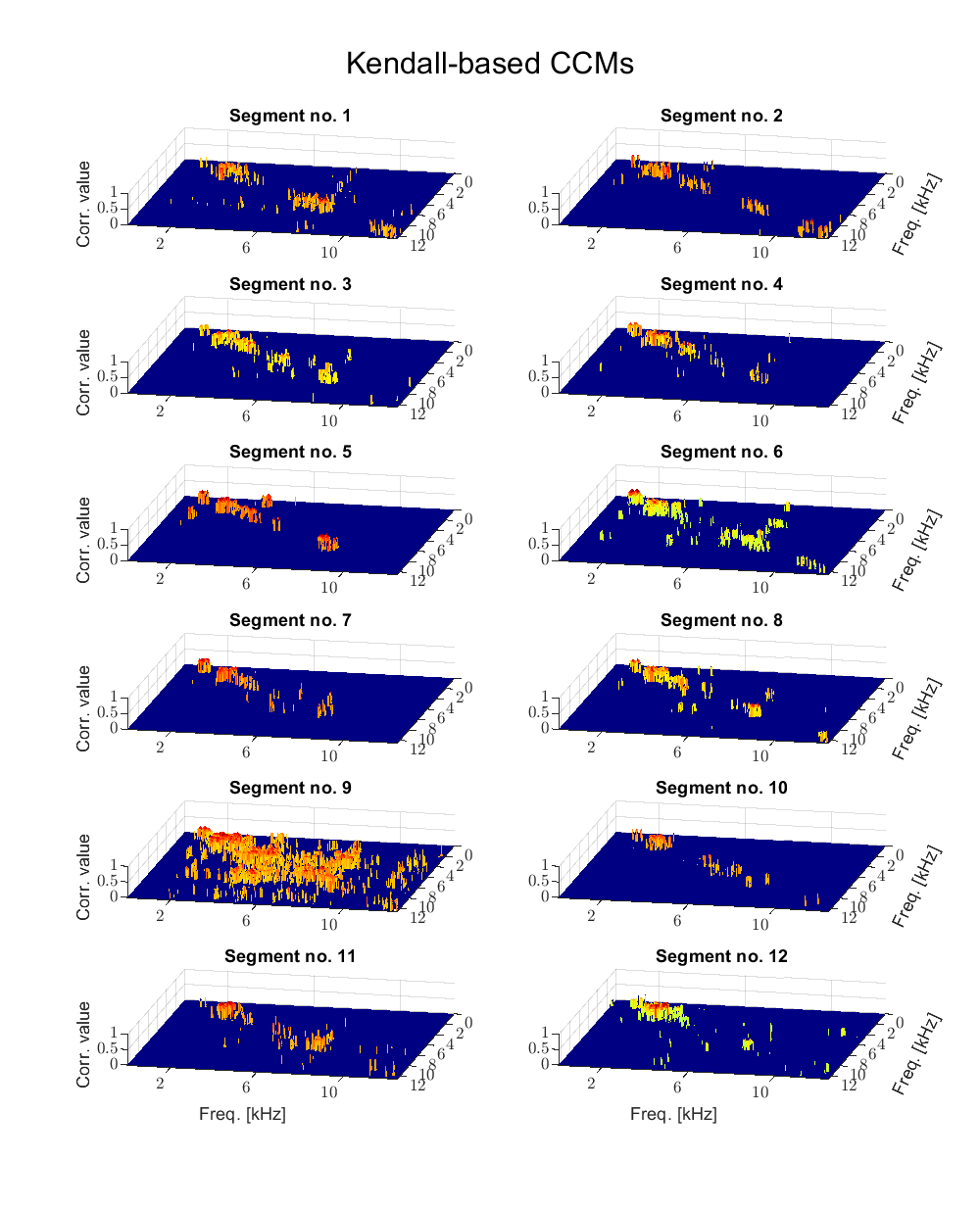}
        \caption{}
    \end{subfigure}%
    ~ 
    \begin{subfigure}[t]{0.5\textwidth}
        \centering
\includegraphics[width=0.99\textwidth]{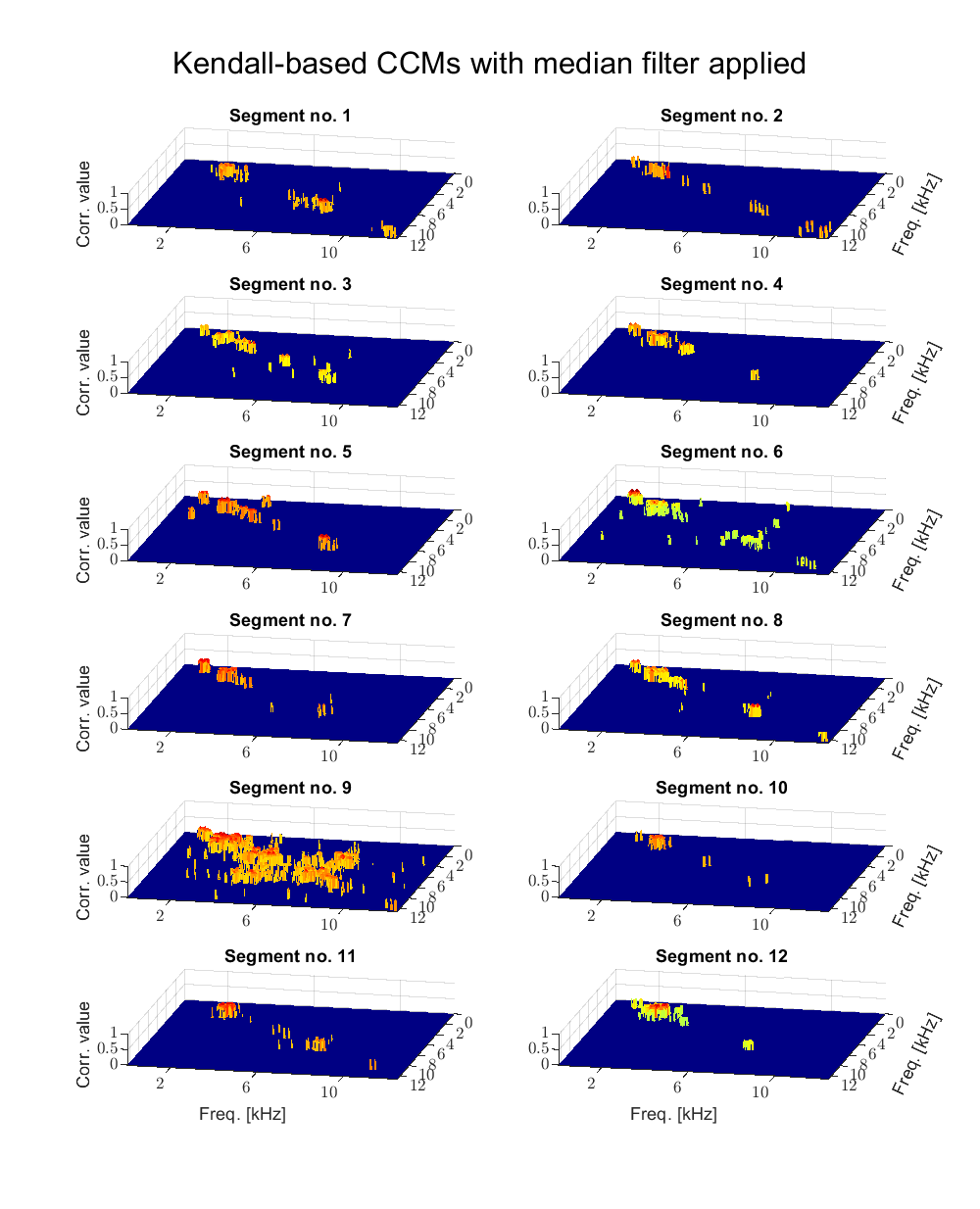}
        \caption{}
    \end{subfigure}
    \caption{Comparison of the results of the proposed correlation measures and median filter utilization i.e. (a) Kendall CMs for each segments of the real vibration signal and (b) corresponding CMs after median filter application.}
    \label{fig:real_maps_ken}
\end{figure}

\begin{figure}[t!]
    \centering
    \begin{subfigure}[t]{0.5\textwidth}
        \centering
\includegraphics[width=0.99\textwidth]{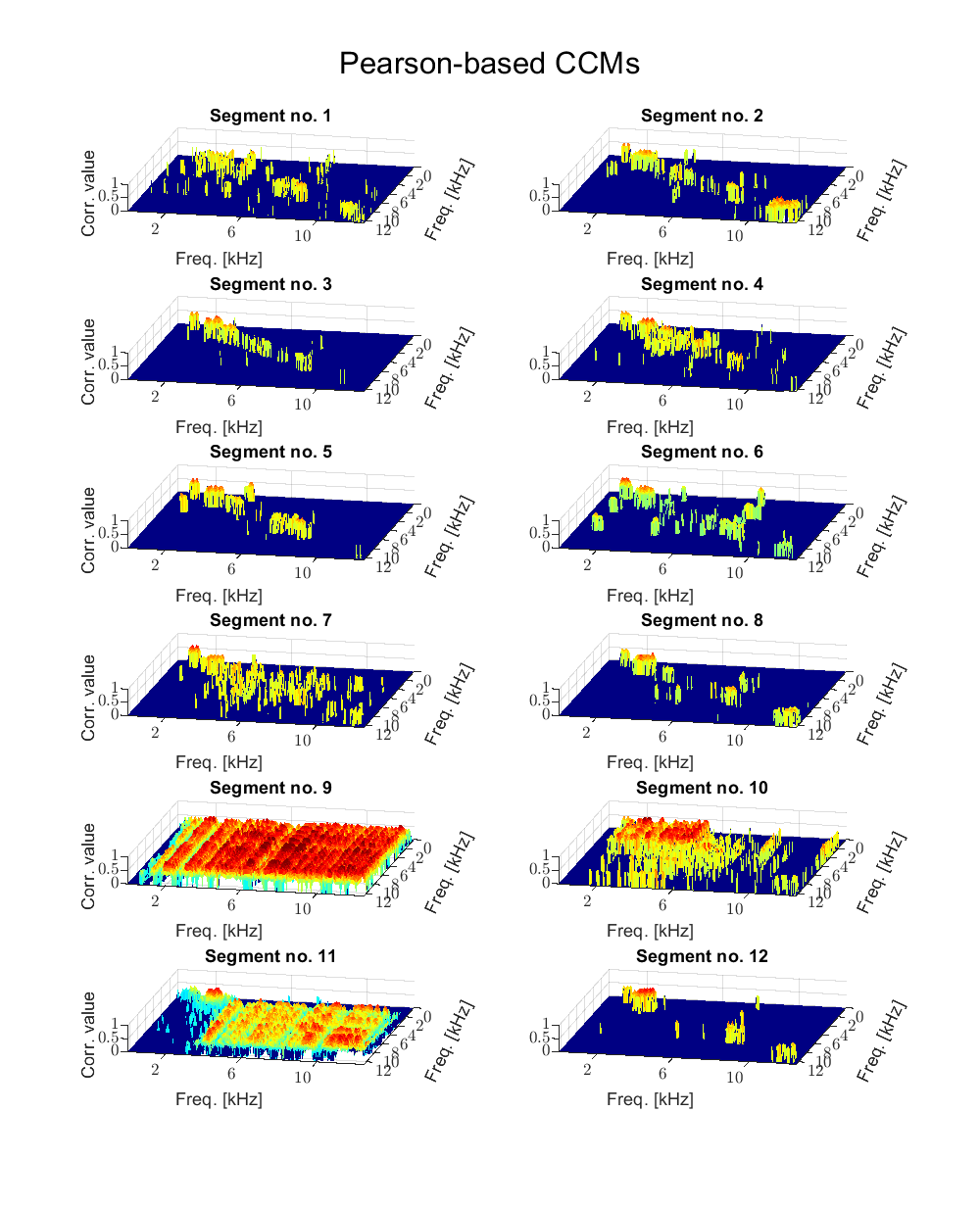}
        \caption{}
    \end{subfigure}%
    ~ 
    \begin{subfigure}[t]{0.5\textwidth}
        \centering
\includegraphics[width=0.99\textwidth]{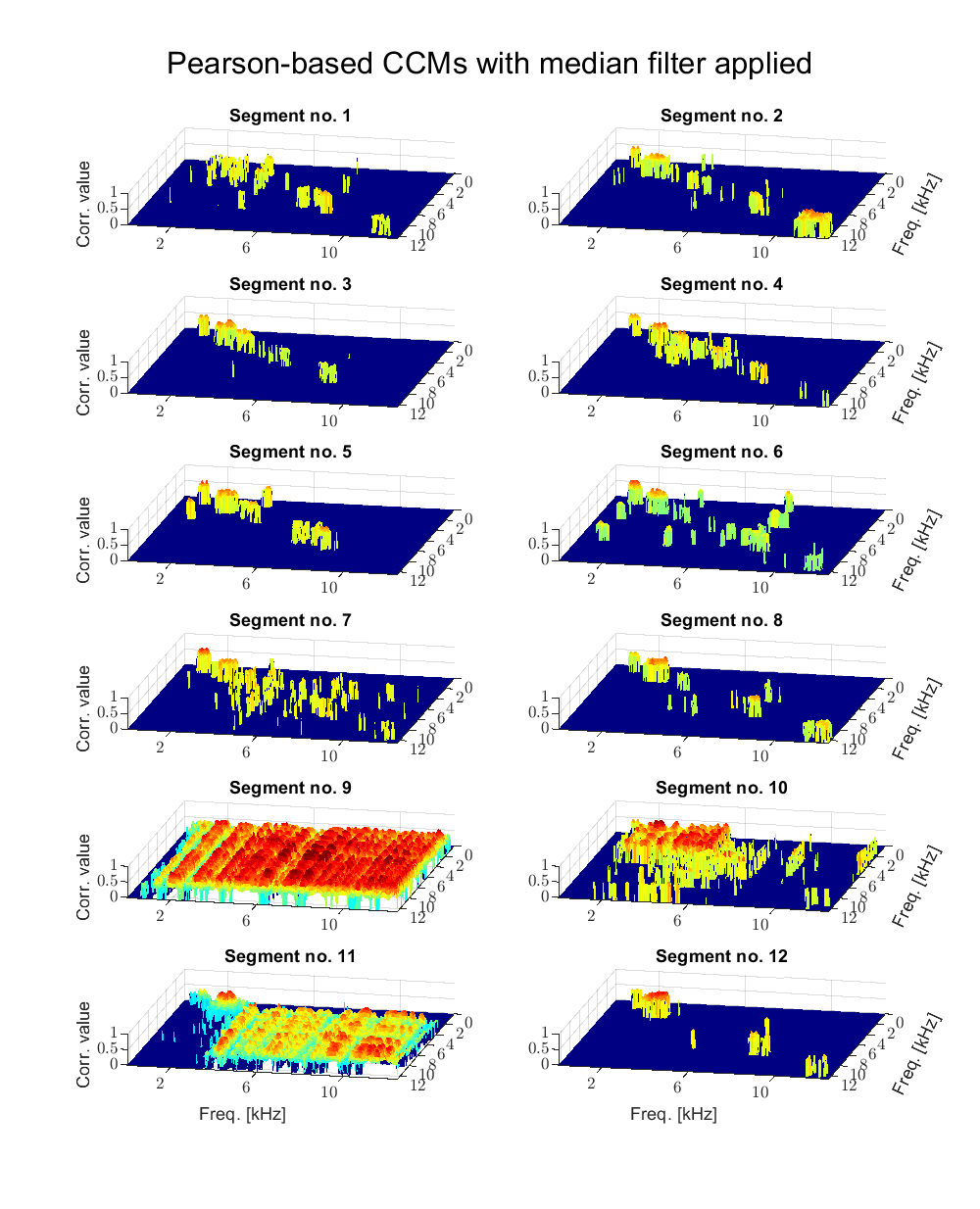}
        \caption{}
    \end{subfigure}
    \caption{Comparison of the results of the proposed correlation measures and median filter utilization i.e. (a) Pearson CMs for each segments of the real vibration signal and (b) corresponding CMs after median filter application.}
    \label{fig:real_maps_pea}
\end{figure}

 Each segment was subjected to the procedure described in Section~\ref{metodo}, including the spectrogram and CM calculation with a median filter applied to finally obtain the CM-based averaged (in the median sense) IFB selector based on Pearson, Kendall, quadrant and trimmed correlation measures. 
 The CMs of each segment without the applied median filter and after the application of the median filter are presented in Fig.~\ref{fig:real_maps_trim} -- Fig.~\ref{fig:real_maps_pea}.

As one can see, depending on the measure of correlation used, the results vary, and it is especially noticeable in segments number 9, 10, and 11. These segments are the part of the signal where non-cyclic impulses appear. In the case of segment number 9, the Pearson CM values are close to 1 almost on the whole map, and it completely covers up the information about the IFB. 
 The application of the median filter smoothed the results when individual spikes appeablack (see segments $1-8$ and 12, Fig.~\ref{fig:real_maps_trim} -- Fig.~\ref{fig:real_maps_pea}). The median filter does not remove/smooth out abnormal high values of correlation if the high values extend across a wide band of frequencies (e.g. see Fig.~\ref{fig:real_maps_pea}, Pearson CM, segment number 9, frequency band $3-12$ kHz). 
 
\begin{figure}[ht!]
  \centering
\includegraphics[width=0.8\textwidth]{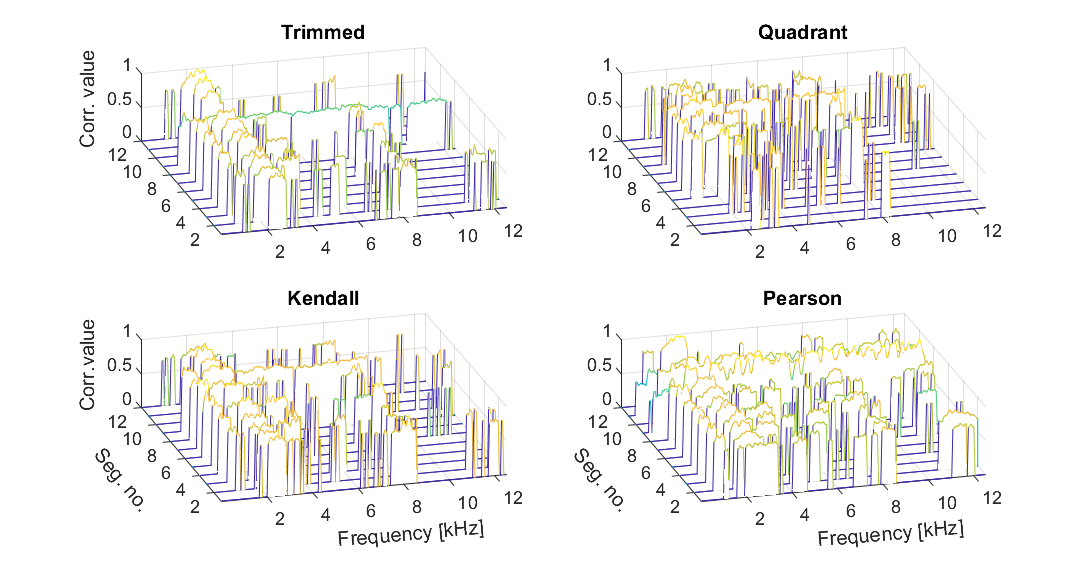}
    \caption{Pearson, Kendall, quadrant, and trimmed CM-based IFB selectors for real vibration signal.}
    \label{fig:sele3}
\end{figure}

\begin{figure}[ht!]
  \centering
\includegraphics[width=0.8\textwidth]{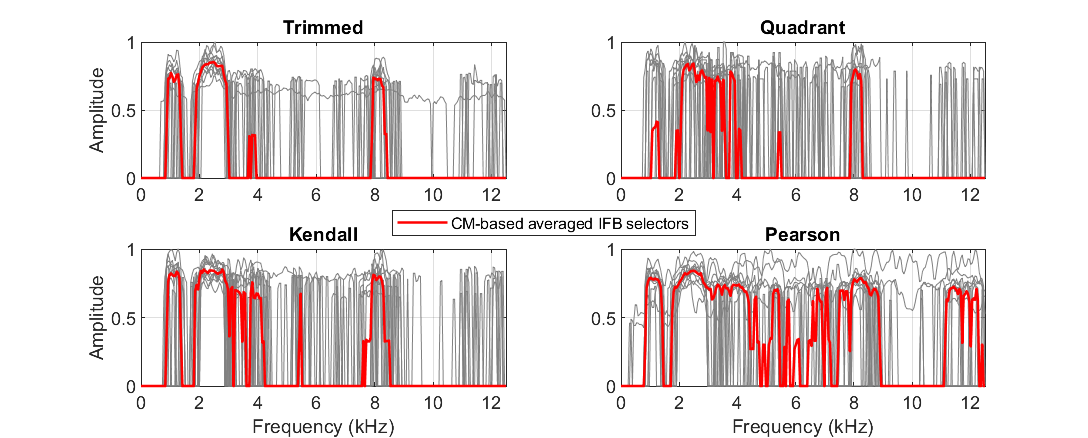}
    \caption{\textcolor{black}{Pearson, Kendall, quadrant, and trimmed CM-based averaged IFB selectors for real vibration signal.}}
    \label{fig:sele_real1}
\end{figure}

 The CM-based IFB selectors for each segment are presented in Fig.~\ref{fig:sele3}. Selectors have different distributions within the frequency range, depending on the measure of the correlation used and the number of segments. The segment number represents the subsequent 0.5 seconds of the signal, and the complexity of the vibration signal in the given time interval varies. The most effective (the best selectivity of the IFB around 2-3 kHz for each segment) seems to be the trimmed and quadrant CM-based IFB selectors. However, there are also high values in some narrow frequency bands that are not related to IFB. The best SNR segment could be selected, but this is a challenge, and the proposed procedure aims to use the information from the entire signal and minimize the impact of outliers. To omit the problem with the selection of the best segment in terms of the SNR all segments are taken into account during the averaged CM-based IFB selector. The median used during the averaging of the IFB selectors of all segments aims to skip extreme values. This property can be observed in Fig.~\ref{fig:sele3} (segment number 9 in each case) and Fig.~\ref{fig:sele_real1} (CM-based averaged IFB selector). The extremely litteblack results of the IFB selector construction for segment number 9 are observed in all IFB selectors consideblack. However, because of the usage of the median during averaging, this result is mostly omitted and does not affect the final result.%
 
 The CM-based averaged IFB selector (calculated from 12 segments) is presented in Fig.~\ref{fig:sele_real1}, marked in the black line (it corresponds to the CM-based averaged IFB selector in the MC simulations, see Section \ref{results}).

\begin{figure*}[ht!]
  \centering
\includegraphics[width=0.85\textwidth]{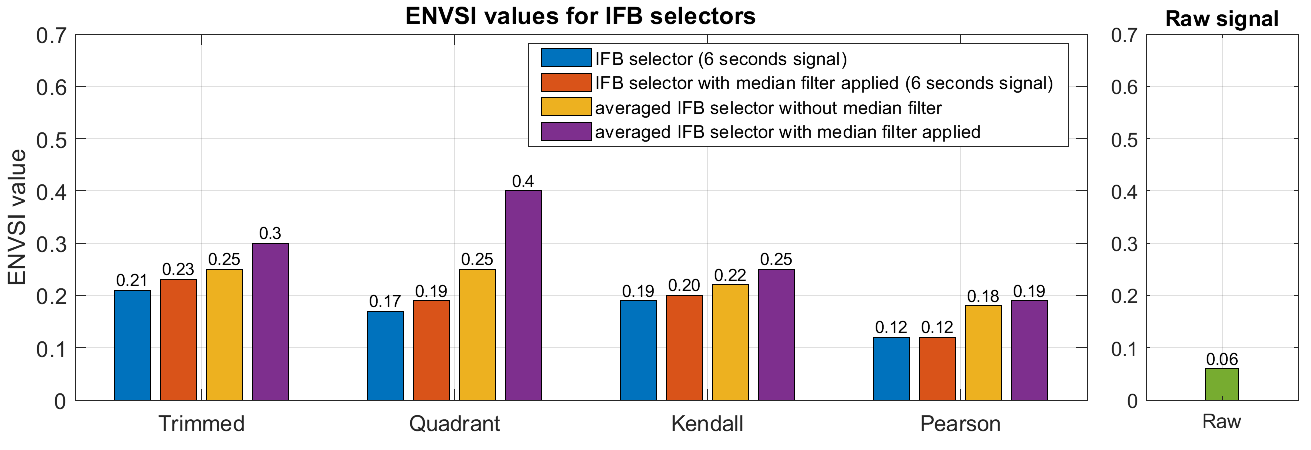}
    \caption{\textcolor{black}{ENVSI values for different IFB selectors for real vibration signal.}}
    \label{tab4}
\end{figure*}

\begin{figure}[ht!]
  \centering
\includegraphics[width=0.8\textwidth]{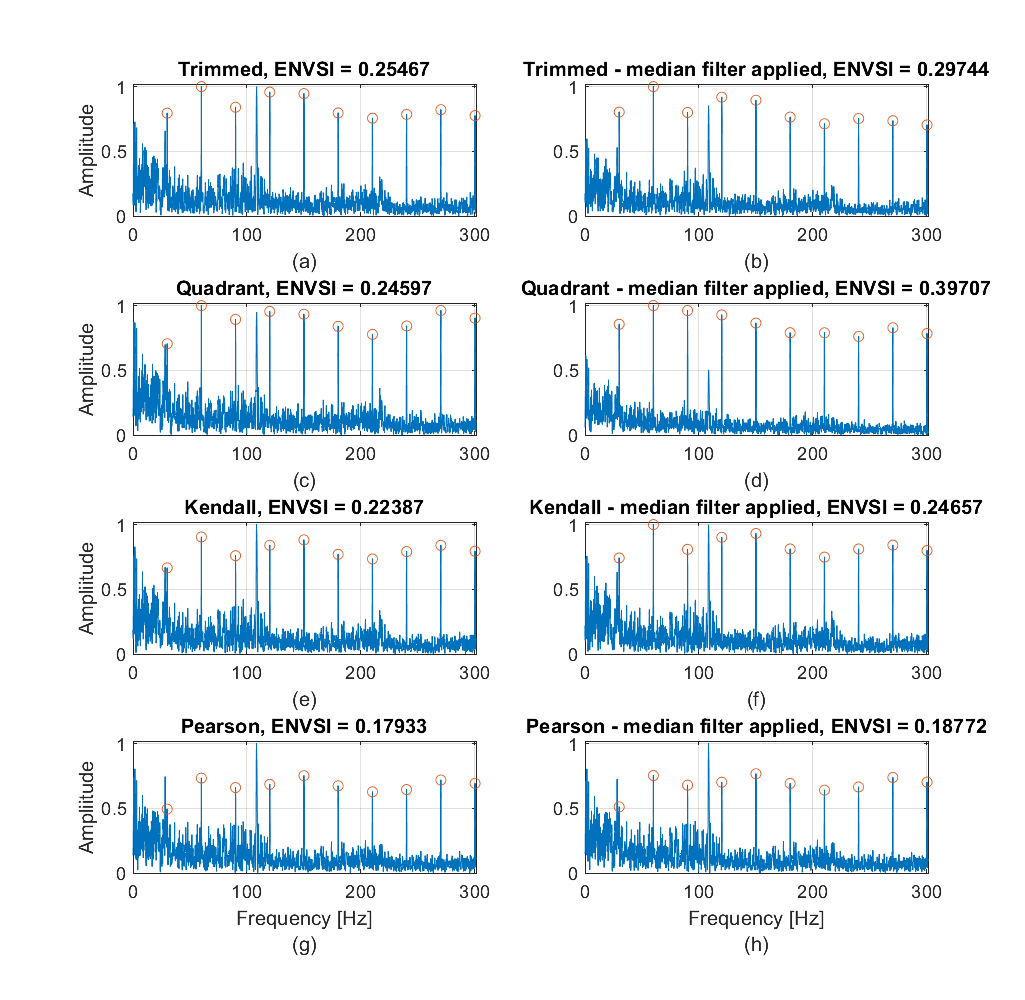}
    \caption{SES of the signal filteblack with the different IFB selectors: a) trimmed IFB selector b) trimmed IFB selector with the median filter applied c) quadrant IFB selector d) quadrant IFB selector with the median filter applied e) Kendall IFB selector f) Kendall IFB selector with the median filter applied g) Pearson IFB selector h) Pearson IFB selector with the median filter applied for the real vibration signal (fault frequency 30 Hz is marked in black circles).}
    \label{fig:envsi_real1}
\end{figure}
The value of ENVSI for each averaged IFB selector is presented in Fig.~\ref{tab4} including the IFB selectors calculated for the overall 6 seconds signal considering the CM with and without the median filter applied. 
The aggregated results that consider the application of new REs in CM, the median utilization of filters, and signal segmentation (see Fig. \ref{tab4}) show that CM-based averaged IFB selectors are effective and can improve SNR and fault detection. Furthermore, the median filter applied in CM improves performance, and additional signal segmentation and averaging of the results using the median can even double the quality of the envelope spectrum of the filteblack signal.
Filtration of the analyzed 6 s length signal with the averaged CM-based IFB selectors has the highest effectiveness. The quality of the envelope spectrum of the filteblack signal is significantly improved and allows for fault frequency identification. 
The best result of the filtration of real vibration signals is achieved by the quadrant CM-based averaged IFB selector with the applied median filter (ENVSI = $0.4$). Compablack to the ENVSI of the raw signal, the improvement in the SES is approximately six times better (improvement by $566\%$). Taking into account the classical Pearson or Kendall approaches, the proposed quadrant CM-based averaged IFB selector provides approximately $233\%$ or (in the case of Kendall CM-based IFB selector) $111\%$ better results. 

The corresponding SES of the averaged IFB selectors are presented in Fig.~~\ref{fig:envsi_real1}, respectively. As one can see, for the consideblack vibration signal, the TCC and QCC performed better results than the PCC and KCC, and the proposed median filter improves the SES of the filteblack signal. In the analyzed signal, the proposed segmentation and median filter application also improve the results of the classical approaches.

\subsection{Comparison with other methods} \label{comparis}
\begin{figure*}[ht!]
  \centering
\includegraphics[width=0.9\textwidth]{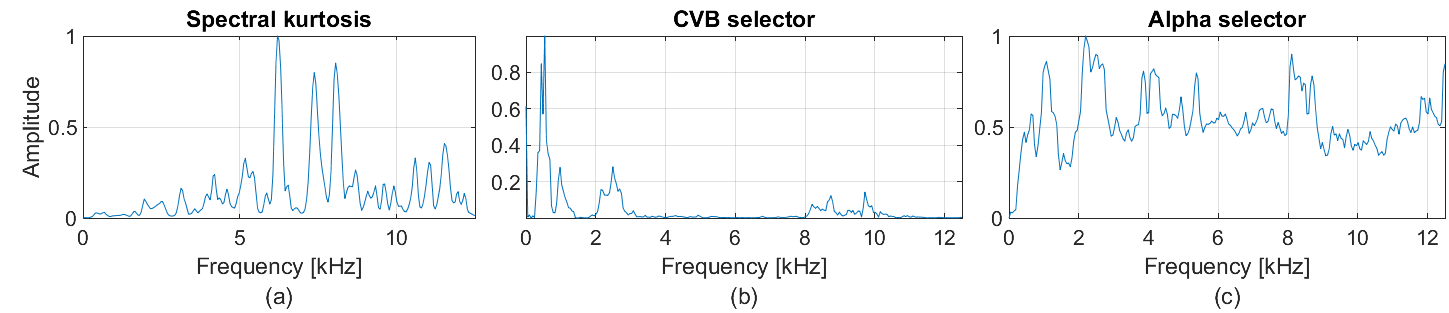}
    \caption{Spectral kurtosis (a), CVB (b) and alpha selectors for real vibration signal.}
    \label{fig:comp_selector_real1}
\end{figure*}

\begin{figure*}[ht!]
  \centering
\includegraphics[width=0.9\textwidth]{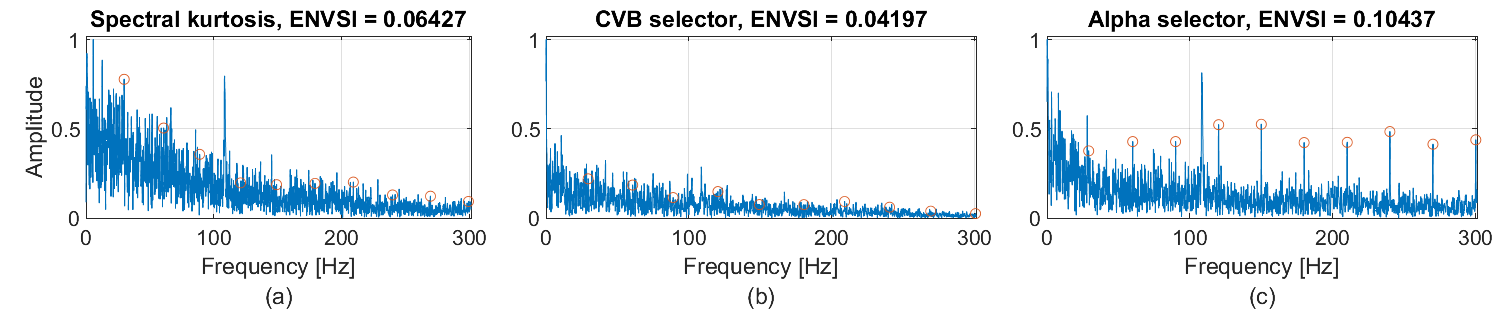}
    \caption{SES of the signals filteblack with spectral kurtosis (a), CVB (b) and alpha (c) selectors \JHSS{with weak fault related harmonics marked in black circles.}}
    \label{fig:comp_envsi_real1}
\end{figure*}

Local damage detection in the presence of non-Gaussian impulsive noise is challenging and requires specific methods. To compare efficiency of the proposed approach, the classical spectral kurtosis is presented, and recently developed approaches developed for an impulsive environment, namely the alpha selector and conditional variance selector (CVB). A more in-depth comparative study can be found in \cite{Hebda-Sobkowicz2020,Hebda-Sobkowicz2020_AS}. As one may see in Fig. \ref{fig:comp_selector_real1}(a), the spectral kurtosis did not identify the informative band as expected. The highest values of normalized (to make comparison with other selectors possible) kurtosis are related to 3 sub-bands between 5-10 kHz. This band is related to non-cyclic impulses recognized as disturbances. Fig.~~\ref{fig:comp_selector_real1}(b) presents the result results of the CVB selector. It is based on the quantiles of the distribution of the consideblack signal and the conditional variance statistic. Details of this procedure are presented in \cite{Hebda-Sobkowicz2020}. Due to the low SNR (impulses of fault are hidden in the background noise), the effectiveness of the CVB selector is limited. There is a visible increase in the selector's value, around 2.5 kHz, but it takes the highest values for low-frequency bands ($0-1.5$ kHz) not related with local fault.
An important diagnostic procedure that indicates the distance between the empirical distribution (corresponding to the signal) and the Gaussian distribution is the alpha selector approach \cite{zak20152987}. The results of the selector for the signal analyzed are presented in Fig.~\ref{fig:comp_selector_real1}(c). It is based on the $\alpha$-stable distribution theory, which in a specific case becomes the Gaussian one. In the signal analyzed with low SNR, the alpha selector outperforms spectral kurtosis and the CVB selector. The values of this selector are greater than 0.5 at almost all frequencies, with maximum values close to 1 focused around $2-2.5$ kHz. As can be seen in Fig.~\ref{fig:comp_envsi_real1}(c), the filtration with the alpha selector improves the SES compablack to the SES of the raw signal (see Fig.\ref{fig:real}, ENVSI $= 0.0631$) and the SES of the signal filteblack with the spectral kurtosis Fig.~\ref{fig:comp_envsi_real1}(a) and the CVB selector Fig.~\ref{fig:comp_envsi_real1}(b). However, the filtration with the alpha selector is around 4 times less effective than it is for the quadrant averaged IFB selector. Even the Pearson averaged IFB selector is almost twice as effective as the filtration with the alpha selector.

\begin{thisnote}
\subsection{Experiment description case 2 - test rig data}    
\label{sec:test_rig2}

The real vibration signal originates from the bearing of the test rig electric motor presented in Fig. \ref{fig:test_rig2}.
\begin{figure}[hbt!]
\centering
\includegraphics[width=0.55\textwidth]{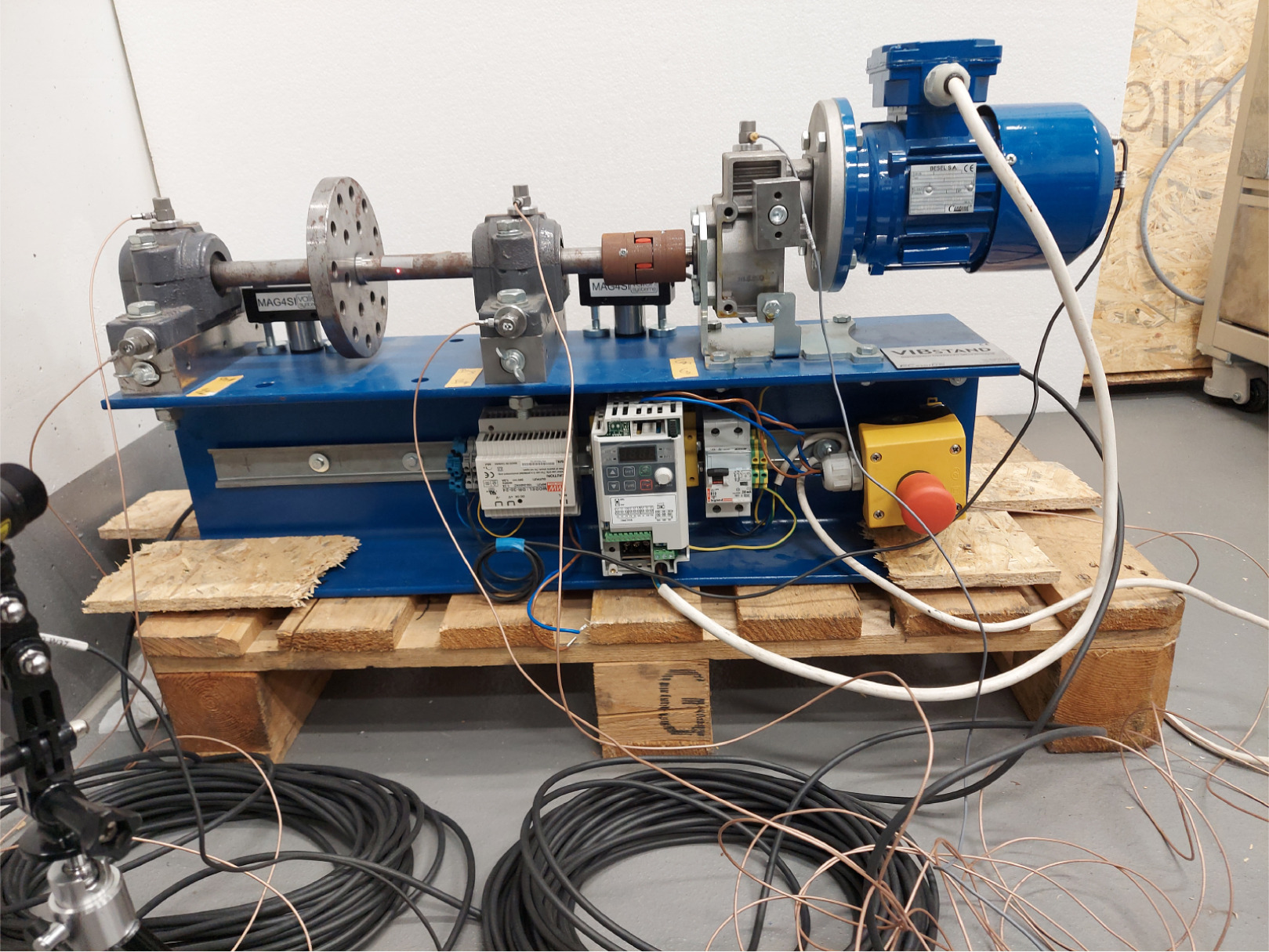}
\caption{\JHSS{Test rig with the electric motor used in the experiment.}}
\label{fig:test_rig2}
\end{figure}
 The signal was acquiblack using a test rig comprising an electric motor, a gearbox, couplings, and two bearings, as shown in Fig. 10. To replicate real-world fault conditions, one of the bearings was deliberately damaged. The signal was recorded at a sampling frequency of 50 kHz using a Bruel \& Kjaer 4189 microphone and a Kistler LabAmp 5165A data acquisition system. During the experiment, the rotational speed was maintained at a non-variable 1041 rpm. The characteristic frequencies of the bearings are listed in Table \ref{tab:freqs_test_rig2}, with the bearings identified as 1205 EKTN9 SKF. One of the bearings exhibited localized damage to the outer race, and the corresponding characteristic frequency for the outer race, highlighted in Table 2, is 91.11 Hz.

\begin{table}[ht!]
  \centering
      \caption{Characteristic frequencies of 1205 EKTN9 bearing}
  \resizebox{0.67\textwidth}{!}{
  \begin{tabular}{|l|l|}
  \hline
     \textbf{Description} & \textbf{Value} \\ \hline
     Rotational frequency of the inner ring & 17.35 Hz\\ \hline
     Rotational freq. of the rolling element and cage assembly & 7 Hz\\ \hline
     Rotational freq. of a rolling element about its own axis & 42.75 Hz\\ \hline
     Over-rolling frequency of one point on the inner ring & 134.44 Hz\\ \hline
     \textbf{Over-rolling frequency of one point on the outer ring} & \textbf{91.11 Hz}\\ \hline
     Over-rolling frequency of one point on a rolling element & 85.5 Hz\\ \hline
  \end{tabular}
  }
  \label{tab:freqs_test_rig2}
\end{table}

\subsection{Test rig data analysis}
\label{sec:real2}
The signal examined in this study is a acoustic signal obtained from bearing of electric motor described in Section \ref{sec:test_rig2}.
The signal has a duration of 6 seconds. The fault frequency is approximately 91 Hz with a central carrier frequency around 22.5 kHz.
The vibration signal and its spectrogram and ES are visible in Fig. \ref{fig:sig_spec_envsi_real2}. Impulses related to
the fault are not visible in the time domain, see Fig. \ref{fig:sig_spec_envsi_real2}(a) nor in the time-frequency domain (i.e. spectrogram), see Fig. \ref{fig:sig_spec_envsi_real2}(b) and nor in the frequency domain in ES, see Fig. \ref{fig:sig_spec_envsi_real2}(c).
\begin{figure}[ht!]
  \centering
    \includegraphics[width=0.7\textwidth]{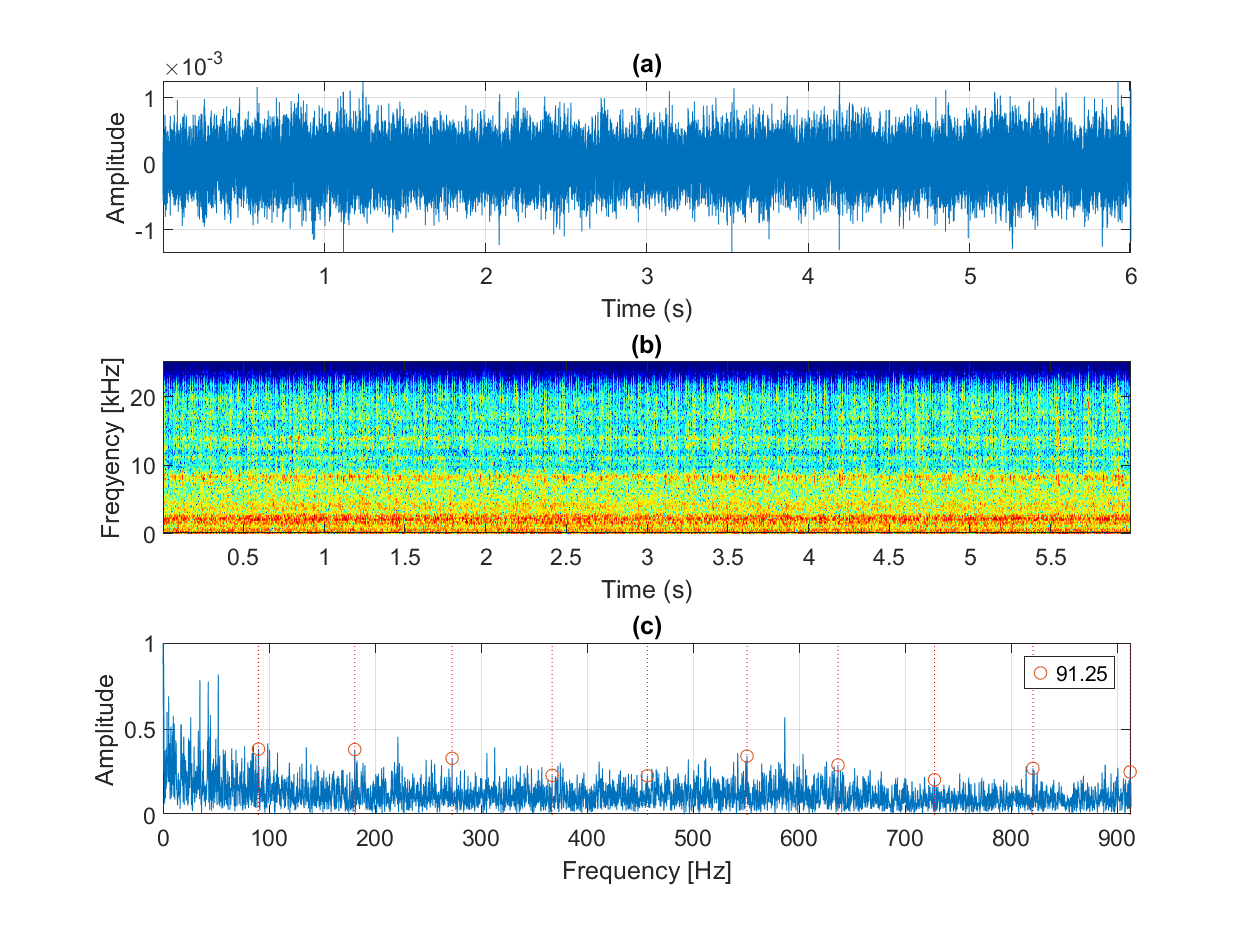}
    \caption{\JHSS{(a) Real vibration signal, (b) its spectrogram and (c) SES with weak fault related harmonics marked in black circles.}}
    \label{fig:sig_spec_envsi_real2}
\end{figure}
In Fig. \ref{fig:real_sel2} the new proposed trimmed and quadrant IFB selectors are presented, as well as, Kendall and Pearson IFB selectors (averaged and with median filter applied). All of them correctly focus on the carrier frequency related with local fault around 22.5 kHz. However some differences are observable, which have the influence during signal filtration.  
\begin{figure}[ht!]
  \centering
\includegraphics[width=0.9\textwidth]{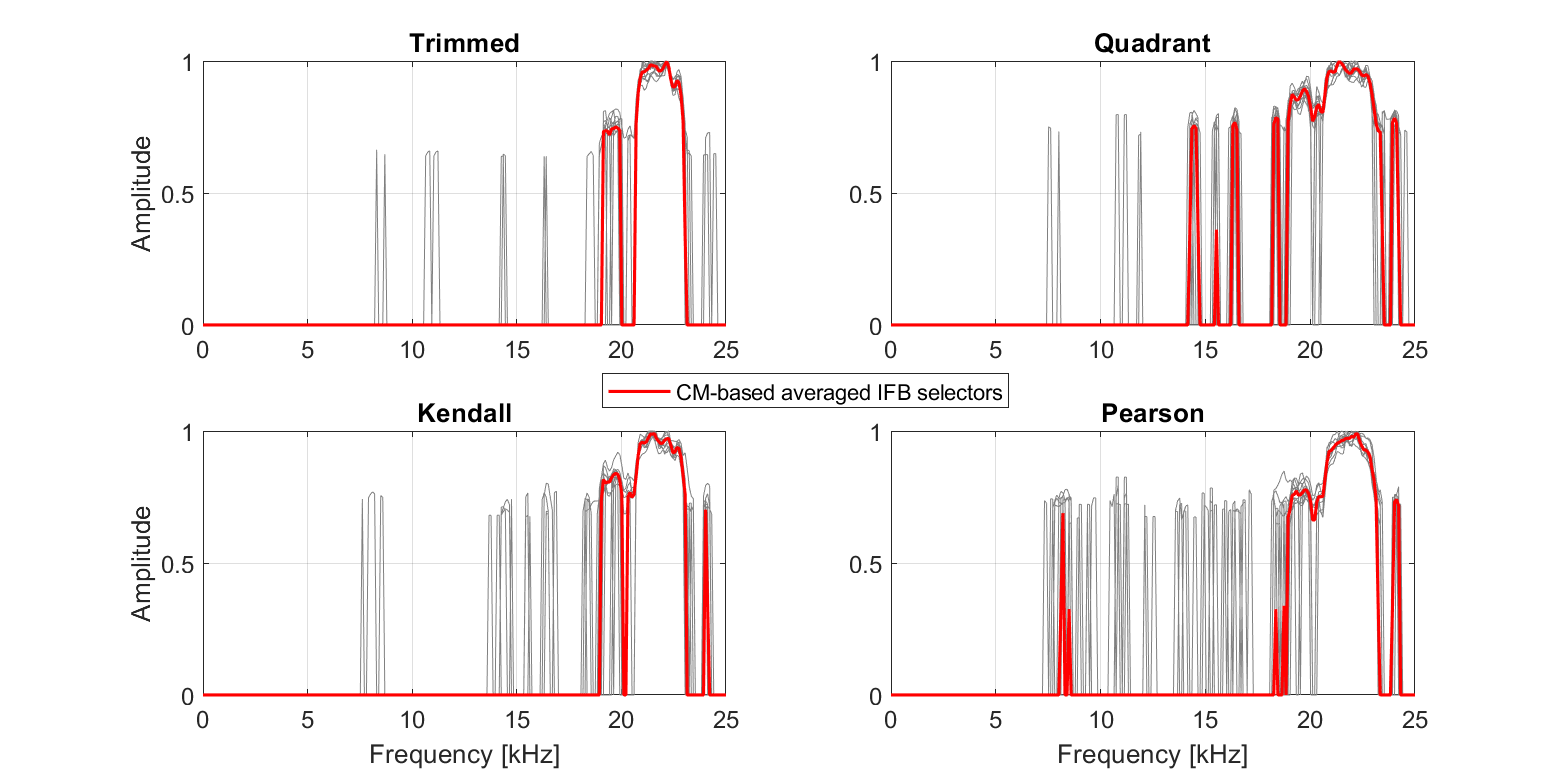}
    \caption{\JHSS{Pearson, Kendall, quadrant, and trimmed CM-based averaged IFB selectors for real acoustic signal.}}
    \label{fig:real_sel2}
\end{figure}
Fig. \ref{fig:envsi_raw2_new_selectors} presents the ES of the filteblack signals with the given selectors. As one can see all IFB selectors effectively filter out the raw signal and after filtration the frequency of fault is clearly visible in the ES.
The best performance is achieved by the Trimmed IFB selector. The ENVSI value takes the highest 0.10 value. 

\begin{figure}[ht!]
  \centering
\includegraphics[width=0.9\textwidth]{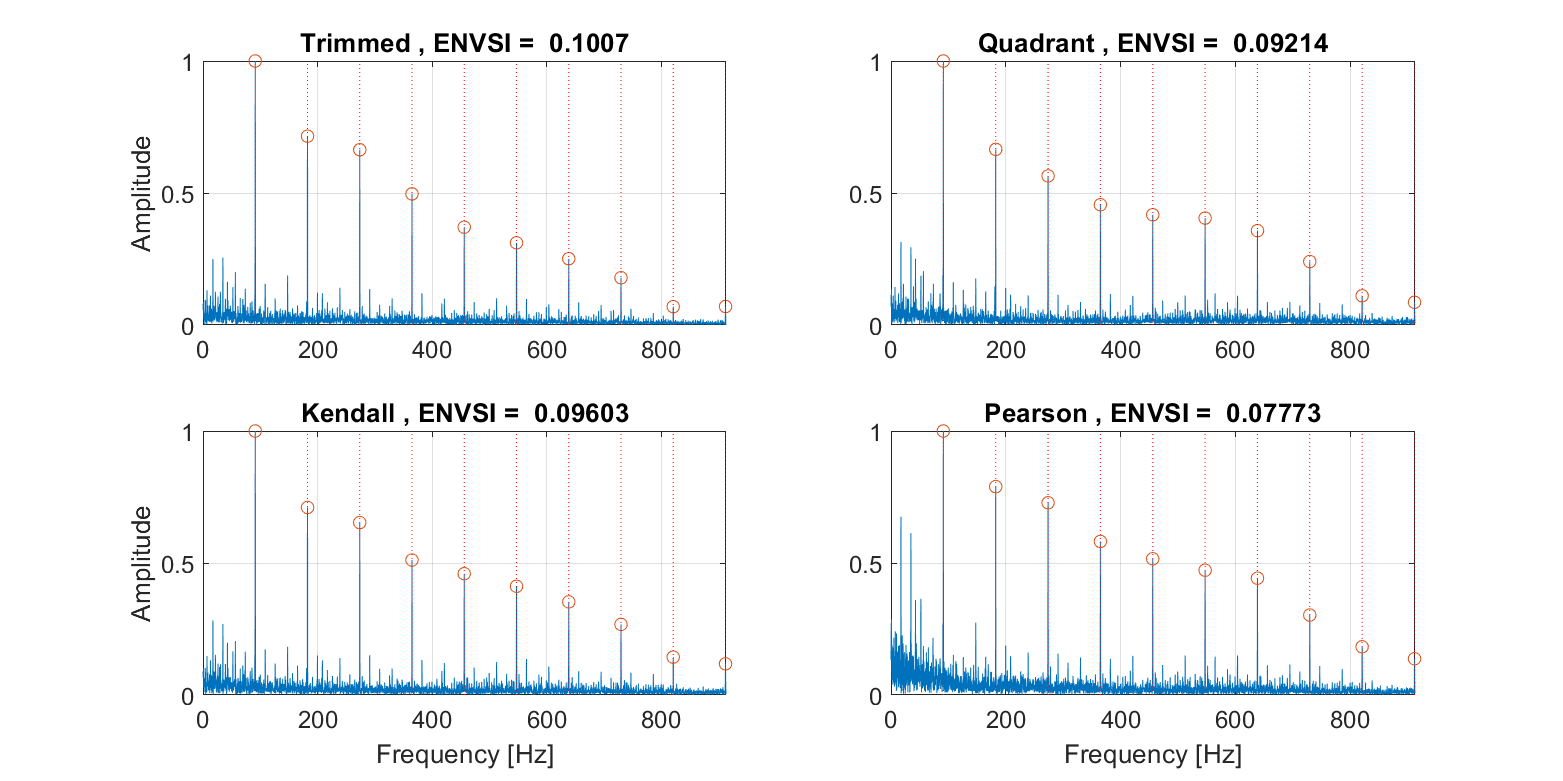}
    \caption{\JHSS{ SES of the signal filteblack with the different IFB selectors: a) trimmed IFB selector with the median filter applied b) quadrant IFB selector with the median filter applied c) Kendall IFB selector d) Pearson IFB selector for the real acoustic signal (fault frequency 91 Hz is marked in black circles).}}
    \label{fig:envsi_raw2_new_selectors}
\end{figure}
In Fig. \ref{fig:raw2_selectors} the results of other known selectors are presented, i.e spectral kurtosis, CVB selector and alpha selector. The effectiveness of spectral kurtosis and CVB is poor in this case as the faulty component is to much hidden in the background noise, only alpha selector takes higher values around IFB but its values also raised on a wide band around 7-20 kHz.   
\begin{figure}[ht!]
  \centering
\includegraphics[width=0.9\textwidth]{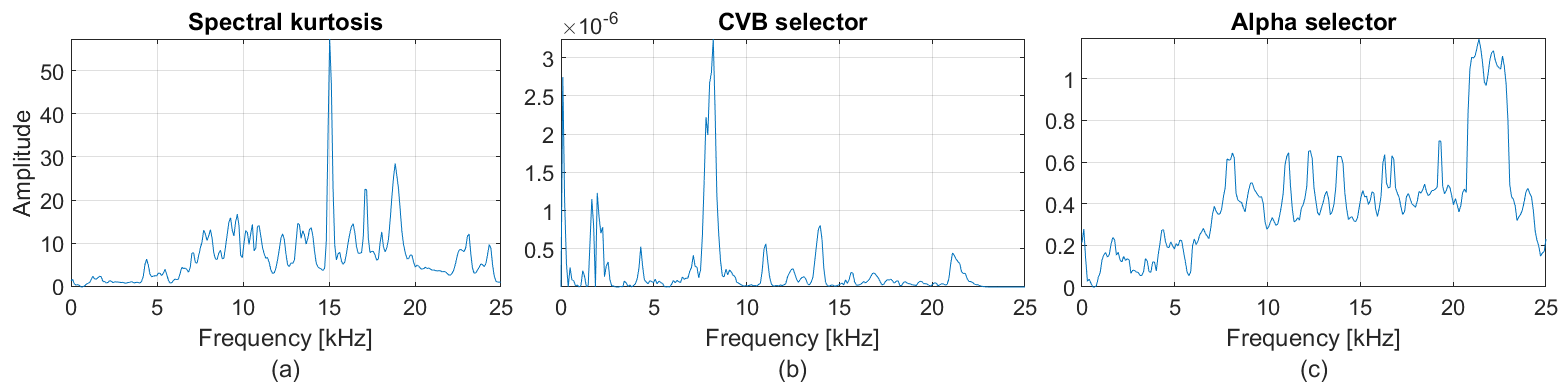}
    \caption{\JHSS{Spectral kurtosis (a), CVB (b) and alpha selectors for real acoustic signal.}}
    \label{fig:raw2_selectors}
\end{figure}
The ES of the filteblack signal with the selectors designed by the spectral kurtosis, CVB and parameter alpha are presented in Fig. \ref{fig:raw2_selectors}.
As one can see, the alpha selector significantly improve the ES comparing to the ES of the raw signal, however the background noise is significantly higher than in case of the filtration with e.g., trimmed IFB selector.
\begin{figure}[ht!]
  \centering
\includegraphics[width=0.9\textwidth]{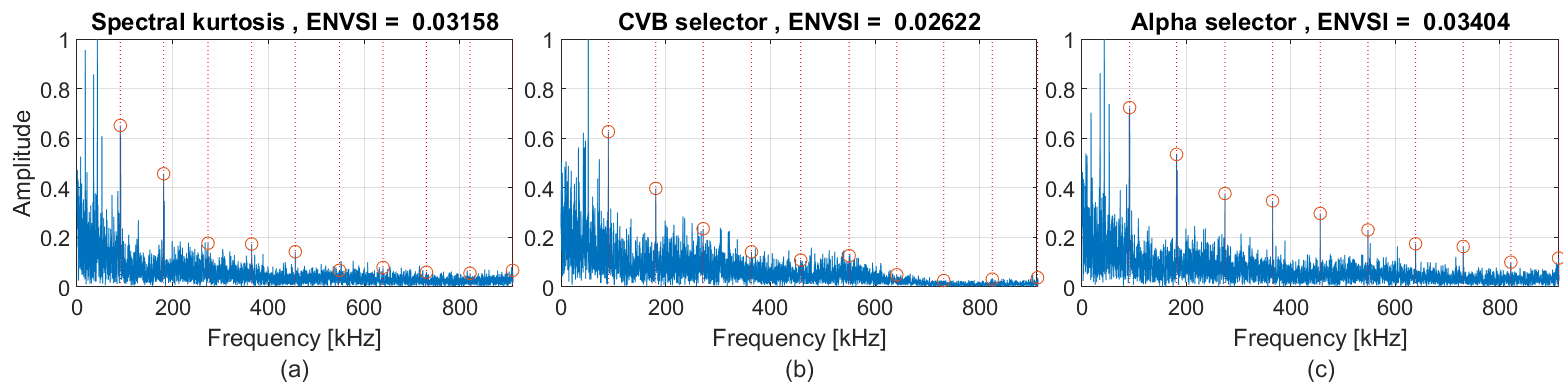}
    \caption{\JHSS{SES of the real acoustic signal filteblack with spectral kurtosis (a), CVB (b) and alpha (c) selectors with weak fault related harmonics marked in black circles.}}
    \label{fig:envsi_raw2_selectors}
\end{figure}
The summary results of ENVSI values of the signals filteblack with the proposed trimmed and quadrant IFB selectors and improved Kendall and Pearson IFB selectors and the results of others selectors known from the literature i.e., spectral kurtosis, CVB and alpha selector are presented in Fig. \ref{fig:envsi_real4}. The best performance for the analyzed acoustic signal is observed with trimmed IFB selector which most effectively filteblack the signal giving the highest value of the ENVSI.
\begin{figure}[ht!]
  \centering
\includegraphics[width=0.85\textwidth]{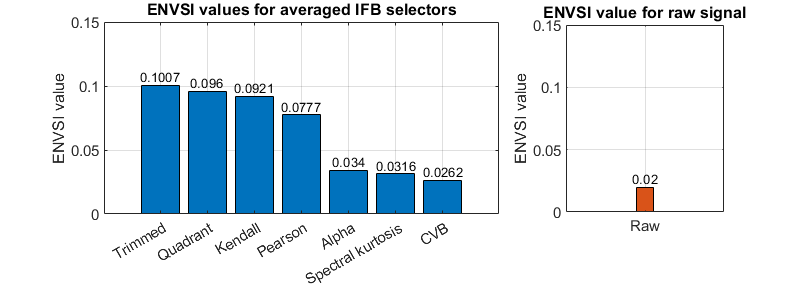}
    \caption{\JHSS{ENVSI values for different IFB selectors for acoustic signal.}}
    \label{fig:envsi_real4}
\end{figure}

\end{thisnote}
\section{Conclusions}
\label{conc}

In the paper, the problem of IFB selection for the detection of local bearing damage was discussed. Vibration signals with non-Gaussian (impulsive) disturbances were consideblack. 
\JHS{The proposed method addresses a kind of matched filter design problem for strongly non-Gaussian signals which combines the signal segmentation, robust correlation measures, spatial median filtering and finally filter amplitude characteristic design. The core of the procedure is to assess the informativeness of the frequency bands, which is used for filter design.}

A bi-frequency map is utilized as a result of correlation for all pairs of subsignals obtained from the spectrogram. Such a map highlights the correlation between subsignals with some temporal structure. To minimize the correlation value between subsignals with non-cyclic impulses, robust measures of correlation \JHSS{i.e. trimmed and quadrant correlation coefficients} are proposed. They are less sensitive to single spikes (outliers).

Simulated data analysis demonstrated that the PCC is highly sensitive to high-amplitude impulsive components. As a result, the Pearson CM emphasizes the non-informative band.
Adjustment of robust correlation measures (TCC and QCC) suppressed the influence of high-amplitude impulses, and the informative frequency band was properly identified. 
Furthermore, the simple technique proposed, i.e. a median filter applied in the CMs additionally improved the results.  Moreover, partitioning the analyzed signal into shorter segment lengths and calculating the CM-based averaged IFB selector (as a median value from selectors of all segments) has the ability to improve the selectivity of the final IFB selector used as a filter characteristic. In addition, the proposed TCC and QCC were compablack with the KCC, commonly known as the robust measure of correlation, and performed better in the case of different parameters of the signal. Furthermore, the computational costs of the CMs calculated by the proposed REs of the correlation are much lower than in the case of the KCC.

\JHSS{The proposed TCC and QCC based approaches performed effectively in IFB selector design for real vibration signals from a crusher in the mining industry and acoustic signals measublack on the test rig. Both approaches correctly emphasize the IFB and suppress other frequency bands. Furthermore, a comparison with established IFB selectors, including spectral kurtosis, the alpha selector, and the CVB selector, demonstrates the superiority of the proposed IFB selector design, which ultimately improves fault frequency detection.}

Based on the analysis presented in this paper, the following conclusions are important to emphasize.
\begin{itemize}
    \item 
In the case of signals with noise close to a Gaussian distribution, the most effective CM in terms of high correlation values for the cyclic impulsive behavior that occurs and in terms of computational complexity is the Pearson CM\JHS{-based selector}. For Gaussian assumptions, the Pearson CM\JHS{-based selector} performs very well, even for low SNR.
\item
When there are non-cyclic impulses in the signal, it is worth considering other REs of correlation. KCC is the most intuitive correlation measure in terms of robustness to outlier values. However, this measure has a high computational complexity that increases 
with the length of the signal. A sensible alternative in terms of computational complexity and robustness to non-cyclic impulses is the trimmed and quadrant CMs proposed in the paper. Based on the results for the 10 second length signal consideblack during simulations, the trimmed CM can be faster by more than 20 times than the computational cost of the Kendall CM, while the quadrant CM can be faster even 40 times than the Kendall CM. However, the quadrant CM seems to be ineffective for small SNRs (even under Gaussian assumptions). The TCC requires some knowledge of the data (outliers number) to properly configure the trimming constant $c$.
\item
The spatial median filter improves CMs that use REs if there are single high correlation values in non-information bands. As a consequence, IFB selectors with spatial median filter applied perform more effective signal filtration. 
\item
 The parameters of the spectrogram, trimming parameter $c$  in the trimmed correlation coefficient, amount and size of segments, as well as the window size of spatial median filter were adopted based on the analysis of real vibration data presented in the paper. The commonly used Hamming window of length 256, with 217 overlapping samples, was applied, and the Fourier transform was calculated for 512 frequency points. Whereas the trimming parameter $c$ in trimmed correlation coefficient was established at $c = 3\%$, according to the literature recommendation [61]. The choice of the parameters could benefit from a more automated procedure - this is an area for future research.
\item
Furthermore, the results of the CM-based IFB selectors indicate that significant improvement in filter characteristic design can be achieved by considering shorter segments of the signal and using the averaged results of the IFB selectors of all segments, i.e., the CM-based averaged (median value) IFB selector for data filtration. The amount and size of segments should be chosen as a compromise between the length of the segment (to achieve the best possible resolution of the signals) and the possibly largest number of segments limited by the exact length of the real vibration signal which is available during analysis. In case of selector averaging the median value is used. A minimum of 5-7 values (signal segments) is typically needed for meaningful results. For the analyzed 6-second signal, 0.5 second segments (which gives 12 values during averaging) were assumed. The sensitivity test of these parameters is the topic of the future research.
\item{Segmentation and median filtering are known techniques in signal processing, however their combined use in our specific context contributes to refining the signal processing pipeline. These techniques help to mitigate noise and improve the quality of the final results without adding unnecessary complexity to the algorithm. Importantly, we believe that the proposed procedure, with its novel application of robust correlation in CMs, addresses a real-world diagnostic challenge in a manner that has not been previously exploblack.}
\item 
Comparison of the proposed methodology with known spectral kurtosis, the alpha selector, and the CVB selector provides significantly better effectiveness during data filtration. For the raw vibration signal consideblack, the spectral kurtosis fails because it is sensitive to impulsive noise. In addition, the CVB selector, which is dedicated to signals with impulsive noise, does not perform effective filtration. It does not sufficiently highlight IFB, which is a consequence of the low SNR in the analyzed signal. Alfa selector performed better than the spectral kurtosis and the CVB selector. However, it is still much worse than the proposed method due to low SNR. Both CVB and alpha selectors investigate the impulsiveness of the SOI, not cyclic behavior.

\end{itemize}

The length of the segment and the number of segments could be a topic for further study, as well as the influence of the number of noncyclic impulses in the signal on the final results of the CMs and IFB selector. 

\JHS{In the paper the influence of the signal parameters is discussed in details (amplitude of SOI, amplitude of non-cyclic impulses, length of the signal), however there is still place to conduct further study and check the influence of other signal parameters like amount of cyclic to non-cyclic impulses.}

\JHS{There is a promising direction for generalization of the proposed algorithm, which involves extending the method to detect arbitrary cyclic or impulsive signals in a variety of challenging non-Gaussian noise environments, including heavy-tailed and impulsive noise. This type of noise, which is commonly encounteblack in real-world diagnostic applications, presents significant challenges to traditional signal processing methods.
Future developments could also explore the incorporation of additional fault types and the ability of the method to deal with a wider range of noise characteristics.}

\section*{Declaration of conflicting interests}

The author(s) declablack no potential conflicts of interest with respect to the research, authorship, and/or publication of this article.

\section*{Acknowledgments}
The work of RZ and AW is supported by the National Center of Science under the Sheng2 project No. UMO-2021/40/Q/ST8/00024 "NonGauMech - New methods of processing non-stationary signals (identification, segmentation, extraction, modeling) with non-Gaussian characteristics to monitor complex mechanical structures".

\bibliography{mybib}

\end{document}